\documentclass[sigconf]{acmart}
\AtBeginDocument{%
  \providecommand\BibTeX{{%
    \normalfont B\kern-0.5em{\scshape i\kern-0.25em b}\kern-0.8em\TeX}}}

\copyrightyear{2020}
\acmYear{2020}
\setcopyright{acmlicensed}
\acmConference[ACSAC 2020]{Annual Computer Security Applications Conference}{December 7--11, 2020}{Austin, USA}
\acmBooktitle{Annual Computer Security Applications Conference (ACSAC 2020), December 7--11, 2020, Austin, USA}
\acmPrice{15.00}
\acmDOI{10.1145/3427228.3427660}
\acmISBN{978-1-4503-8858-0/20/12}
\usepackage{acmart-taps}
\usepackage{booktabs} % For formal tables
\usepackage{neuralnetwork}
\usepackage{algorithm}
\usepackage{algpseudocode}

\usepackage{subfigure}
\usepackage{amsmath}
\usepackage{tabularx}
\usepackage{multirow}
\usepackage{comment}

\usepackage{wasysym}%
\usepackage{pifont}%
\newcommand{\cmark}{\ding{51}}%
\newcommand{\xmark}{\ding{55}}%
\newcommand{\tDot}{$\CIRCLE$}

\newcommand{\thdot}{$\LEFTcircle$}
\usepackage{array}
\usepackage{graphicx}
\usepackage{multirow}

\usepackage[makeroom]{cancel}

\usepackage{xspace}
\newcommand{\wadi}{WADI Testbed\xspace}

\newcommand{\Paragraphs}[1]{\smallskip\noindent{\textbf{\emph{ #1}}}}

\newcommand{\nodetextx}[2]{\ifnum #2=5 $r_{n}$ \else $r_#2$ \fi}
\newcommand{\nodetexty}[2]{\ifnum #2=5 $v_{n}$ \else $v_#2$ \fi}
\newcommand{\hiddentext}[2]{\ifnum #2=4 $h_{p}$ \else $h_#2$ \fi}

\newcommand\AND{\textbf{\&\&}}
\newcommand{\whitebox}{white box\xspace}
\newcommand{\blackbox}{black box\xspace}
\newcommand{\approachI}{iterative\xspace}
\newcommand{\approachII}{learning based\xspace}
\newcommand{\smean}{\ensuremath{\mu_{\bar{x}}}}
\newcommand{\sstd}{\ensuremath{\sigma_{\bar{x}}}}

\acmBadgeR[https://www.acsac.org/2020/submissions/papers/artifacts/]{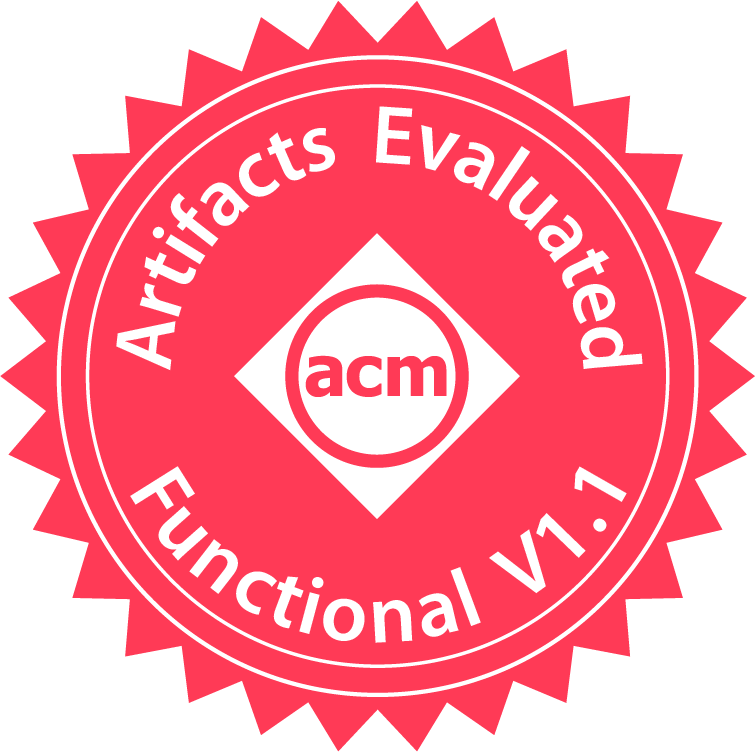}

\begin{document}

\title[Constrained Concealment Attacks against Reconstruction-based detectors in ICS]{Constrained Concealment Attacks against Reconstruction-based Anomaly Detectors in Industrial Control Systems}

\author{Alessandro Erba}
\email{alessandro.erba@cispa.saarland}
\affiliation{%
  \institution{CISPA Helmholtz Center for Information Security}
  \streetaddress{Stuhlsatzenhaus 5, 66123 Saarbr\"ucken}
  \city{Saarbr\"ucken}
  \state{Germany}
  \postcode{66123}
}
\additionalaffiliation{%
\institution{Saarbr\"ucken Graduate School of Computer Science, Saarland University}}
\authornote{A part of this work was done while Alessandro Erba was student at Politecnico di Milano, visiting SUTD.}

\author{Riccardo Taormina}
\email{r.taormina@tudelft.nl}
\affiliation{%
  \institution{Delft University of Technology}
  \streetaddress{Mekelweg 5}
  \city{Delft}
  \state{Netherlands}
  \postcode{ 2628 CD }
}
\author{Stefano Galelli}
\email{stefano_galelli@sutd.edu.sg}
\orcid{1234-5678-9012}
\affiliation{%
  \institution{Singapore University of Technology and Design}
  \streetaddress{8 Somapah Rd}
  \city{Singapore}
  \postcode{487372}
}
\author{Marcello Pogliani}
\email{marcello.pogliani@polimi.it}
\affiliation{%
  \institution{Politecnico di Milano}
  \streetaddress{Piazza Leonardo da Vinci 32}
  \city{Milan}
  \state{Italy}
  \postcode{20133}
}

\author{Michele Carminati}
\email{michele.carminati@polimi.it}
\affiliation{%
  \institution{Politecnico di Milano}
  \streetaddress{Piazza Leonardo da Vinci 32}
  \city{Milan}
  \state{Italy}
  \postcode{20133}
}

\author{Stefano Zanero}
\email{stefano.zanero@polimi.it}
\affiliation{%
  \institution{Politecnico di Milano}
  \streetaddress{Piazza Leonardo da Vinci 32}
  \city{Milan}
  \state{Italy}
  \postcode{20133}
}

\author{Nils Ole Tippenhauer}
\email{tippenhauer@cispa.saarland}
\affiliation{%
  \institution{CISPA Helmholtz Center for Information Security}
  \streetaddress{Stuhlsatzenhaus 5, 66123 Saarbr\"cken}
  \city{Saarbr\"ucken}
  \state{Germany}
  \postcode{66123}
}

\begin{abstract}
Recently, reconstruction-based anomaly detection was proposed as an effective technique to detect attacks in dynamic industrial control networks. Unlike classical network anomaly detectors that observe the network traffic, reconstruction-based detectors operate on the measured sensor data, leveraging physical process models learned a priori. 

In this work, we investigate different approaches to evade prior-work reconstruction-based anomaly detectors by manipulating sensor data so that the attack is concealed. We find that replay attacks (commonly assumed to be very strong) show bad performance (i.e., increasing the number of alarms) if the attacker is constrained to manipulate less than 95\% of all features in the system, as hidden correlations between the features are not replicated well. To address this, we propose two novel attacks that manipulate a subset of the sensor readings, leveraging learned physical constraints of the system. Our attacks feature two different attacker models: A \whitebox attacker, which uses an optimization approach with a detection oracle, and a \blackbox attacker, which uses an autoencoder to translate anomalous data into normal data.
We evaluate our implementation on two different datasets from the water distribution domain, showing that the detector's Recall drops from 0.68 to 0.12 by manipulating 4 sensors out of 82 in WADI dataset. In addition,  we show that our \blackbox attacks are transferable to different detectors: They work against autoencoder-, LSTM-, and CNN-based detectors. Finally,  we implement and demonstrate our attacks on a real industrial testbed to demonstrate their feasibility in real-time.
\end{abstract}




\renewcommand{\shortauthors}{A. Erba et al.}

\maketitle

\section{Introduction}% 1-1.5 pages
Computational and physical infrastructures are nowadays interconnected. Computers, communication networks, sensors, and actuators allow to control physical processes, resulting in what is known as cyber-physical systems (CPS). Examples of such systems are interconnected critical industrial control systems (ICS) like power grids~\cite{liu2011false}, water supply systems~\cite{AminLitricoSastryBayen2013PartI}, and autonomous vehicles~\cite{cardenasAminLinHuangHuangSastry}. 

The integration of modern security features into existing ICS is challenging, as solutions have to be backward compatible with decades-old devices in the field, which do not support authentication or encryption. As a result, attackers with the goal of damaging the process and local access to the network are usually assumed to be able to eavesdrop on traffic, send malicious commands to actuators, and spoof sensor values to hide problems in the physical process~\cite{huitsing2008attack,weinbergerStuxnet, garcia17hey}.
Such activities produce anomalies in the physical sensor data that can be successfully leveraged for attack detection. Hence, attackers can attempt to conceal the physical anomalies through replay attacks~\cite{mo2009secure} or through stealthy attacks based on solving models of the (known)  physical processes~\cite{teixeira2012revealing,urbina16limiting}.

A promising anomaly detection technique in industrial control systems involves the use of machine learning-based classifiers and, in particular, reconstruction-based classifiers as proposed in~\cite{taormina2018deep, goh2017anomaly, kravchik2018detecting}. An attacker who wants to conceal the physical anomalies from this detector will modify a sample to induce a wrong classification outcome: This can be framed as an Adversarial Machine Learning (AML) evasion attack. So far, no systematic analysis of evasion attack against reconstruction-based detectors has been\break proposed.

Evasion attacks in the context of ICS (which we call \emph{Concealment Attacks} to distinguish them from the general case) face novel and specific challenges, which make standard AML techniques~\cite{huang2011adversarial} not directly applicable.

In particular, adversarial examples\footnote{We differentiate between \emph{sample} (original set of sensor readings), and \emph{adversarial example} (manipulated set of sensor readings).} obtained with a concealment attack must meet four requirements.
\textbf{R1}: Due to the distributed nature of the system, the attacker is constrained to manipulate only a subset of features.
\textbf{R2}: Adversarial examples must meet the temporal and spatial correlations expected from the observed physical processes~\cite{taormina18battle,hayes2015contextual}. Particularly, adversarial examples must not introduce contextual anomalies (i.e., observations classified as abnormal only when viewed against other variables that characterize the behavior of the physical process~\cite{hayes2015contextual}).
\textbf{R3}: Most AML attacks in other domains target end-to-end Neural Network Classifiers instead of reconstruction-based classifiers. We realized that this requires the attacker to optimize the Mean Squared Error loss instead of optimizing the cross-entropy loss (Section~\ref{sec:positioning}). 
\textbf{R4}: To the best of our knowledge, previous work either assumes unlimited computational power to compute ideal pertubations~\cite{carlini2019evaluating} or assumes static systems that allow universal adversarial perturbations~\cite{moosavi2017universal, li2019adversarial}. For real-time attacks\footnote{With real-time, we mean examples are crafted w.r.t. the current dynamic state of the system, in less time than the sampling rate (e.g., 10ms).} on dynamic ICS, neither approach is feasible, and new solutions are required.

In this work, \emph{we propose and evaluate constrained concealment attacks against reconstruction-based anomaly detectors}. To meet R1, we formalize a detailed attacker model and evaluate different settings relating to attacker constraints, i.e., the number of features under the control of the attacker. To meet R2, the attacker leverages passive observation of system behavior to approximate how realistic examples should behave. Based on that, we consider a \whitebox attacker that can leverage knowledge on the system to perform iterative attacks on general reconstruction-based detectors (meeting R3). Moreover, we consider a  \blackbox attacker and show that it is possible to craft effective adversarial samples in milliseconds (meeting R3 and R4). Our implementation meets all four requirements for Reconstruction-based detectors.

We summarize our main contributions as follows: 
\begin{itemize}
    \item We propose a detailed attacker model that formalizes implicit models in prior work, introduce constraints on the attacker motivated by real-world ICS, and provide an AML taxonomy for the attacker.
    \item We show that replay attacks do not conceal anomalies when the attacker is constrained to manipulate less than 95\% of the ICS features, due to physical correlations that are not exploited by such attacks.
    \item We propose and design concealment attacks on ICS process-based anomaly detectors which produce examples that do not violate correlations (outperforming replay attacks in constrained scenarios). A \whitebox attacker exploits the knowledge of the Anomaly Detection System launching an iterative attack based on coordinate descent algorithm. A \blackbox attacker without the knowledge of the Anomaly Detection System uses learning-based attack leveraging adversarially trained autoencoders\footnote{Note: Not to be confused with Adversarial Autoencoders~\cite{makhzani2015adversarial}.}.%, enabling dynamic attacks in real-time.
    \item We evaluate and discuss the proposed attacks, and compare their performance against replay attacks. The evaluation is conducted over a simulated ICS process dataset and a real ICS process dataset, both containing data of water distribution systems\footnote{Implementation available at \url{https://github.com/scy-phy/ICS-Evasion-Attacks}}.
    \item We practically implement and demonstrate the attacks in real-word Industrial Control System testbed, and show that they are possible in real-time.
\end{itemize}

The remainder of this work is structured as follows. Background is introduced in Section~\ref{sec:background}. We present the problem of adversarial learning attacks on ML-based detectors in Section~\ref{sec:problem}. Our design  is proposed in Section~\ref{sec:design}, and its implementation and evaluation is presented in Section~\ref{sec:implementation}. We summarize related work in Section~\ref{sec:related}. The paper is concluded in Section~\ref{sec:conclusions}.

\section{Background}\label{sec:background} % 1-2 pages
In this section, we provide a brief overview on Industrial Control Systems and Evasion Attacks. %, and **Attack detection in ICS**
A review of related work is presented in Section~\ref{sec:related}. 
\subsection{Industrial Control Systems}
Industrial Control Systems are universally employed to control an industrial process~\cite{giraldo18survey}. 
%While the physical components actuate and measure the process, the cyber components control the process stages. \
In an ICS, \emph{Physical} components include the hardware equipment required to execute the process; among these, actuators and sensors represent the junction point between \emph{Cyber} and \emph{Physical} components. 
%For example, in a water facility, water tanks, pipes, pumps, and flow meters can be found.
\emph{Cyber} components comprise the computer hardware and software that is deployed to execute the plant control logic and monitoring. Typical industrial control hardware contain Programmable Logic Controllers (PLCs) and a Supervisory Control and Data Acquisition (SCADA) system. %, Human-Machine Interface (HMI). 
In an ICS, one or more PLCs implement the process control logic by monitoring sensor values and sending commands to actuators. Sensor values and actuator states are reported from PLC to the SCADA system.  
In a distributed ICS, several PLCs control the system; taking control of a sub-process governed by a single PLC can disrupt the system. 
Attacks targeting both the cyber and physical components of industrial processes have occurred during the past decades. A notable example is Stuxnet~\cite{weinbergerStuxnet}, an attack that targeted the physical part of an ICS to reduce rotation frequencies of nuclear centrifuges, and the cyber component to spoof reported sensor readings (via a man-in-the-middle attack) thus avoiding anomaly\break detection.

\subsection{Evasion Attacks}

In AML, an evasion attack is launched by an adversary to control the output behavior of a machine learning model through crafted inputs i.e.,~\emph{adversarial examples}. Several evasion attack and defense mechanisms have been proposed in the context of image processing~\cite{szegedy13intriguing}, speech recognition~\cite{carlini2018audio} and malware detection~\cite{grosse17adversarial}.

The authors of~\cite{biggio18wild} characterize attacks on machine learning models using a 4-tuple representation of the attackers' knowledge of the system under attack, the training dataset $\mathcal{D}$, the feature set $\mathcal{X}$, the learning algorithm $f$, and the trained parameters $w$.
In an adversarial setting, an attacker has complete or partial knowledge of components; partial knowledge of a component is denoted with the symbols $\hat{\mathcal{D}}$, $\hat{\mathcal{X}}$, $\hat{f}$ and $\hat{w}$ respectively.
In particular, the authors characterize three types of attack scenarios:
Perfect-knowledge \whitebox attackers $(\mathcal{D}, \mathcal{X}, f, w)$; Limited-knowledge gray box attacks %with
$(\hat{\mathcal{D}}, \mathcal{X}, f, \hat{w})$; Zero-knowledge \blackbox attacks $(\hat{\mathcal{D}}, \hat{\mathcal{X}}, \hat{f}, \hat{w})$. Attacks are achieved by solving an optimization problem that minimizes distance between the sample and the adversarial example e.g. by minimizing norms: \emph{L\textsubscript{0}}, \emph{L\textsubscript{2}}, \emph{L\textsubscript{$\infty$}}. In Section~\ref{sec:problem}, we use this notation to introduce our proposed solution and position it within the related literature.

\begin{figure}
    %\begin{center}
    \includegraphics[width=0.7\linewidth]{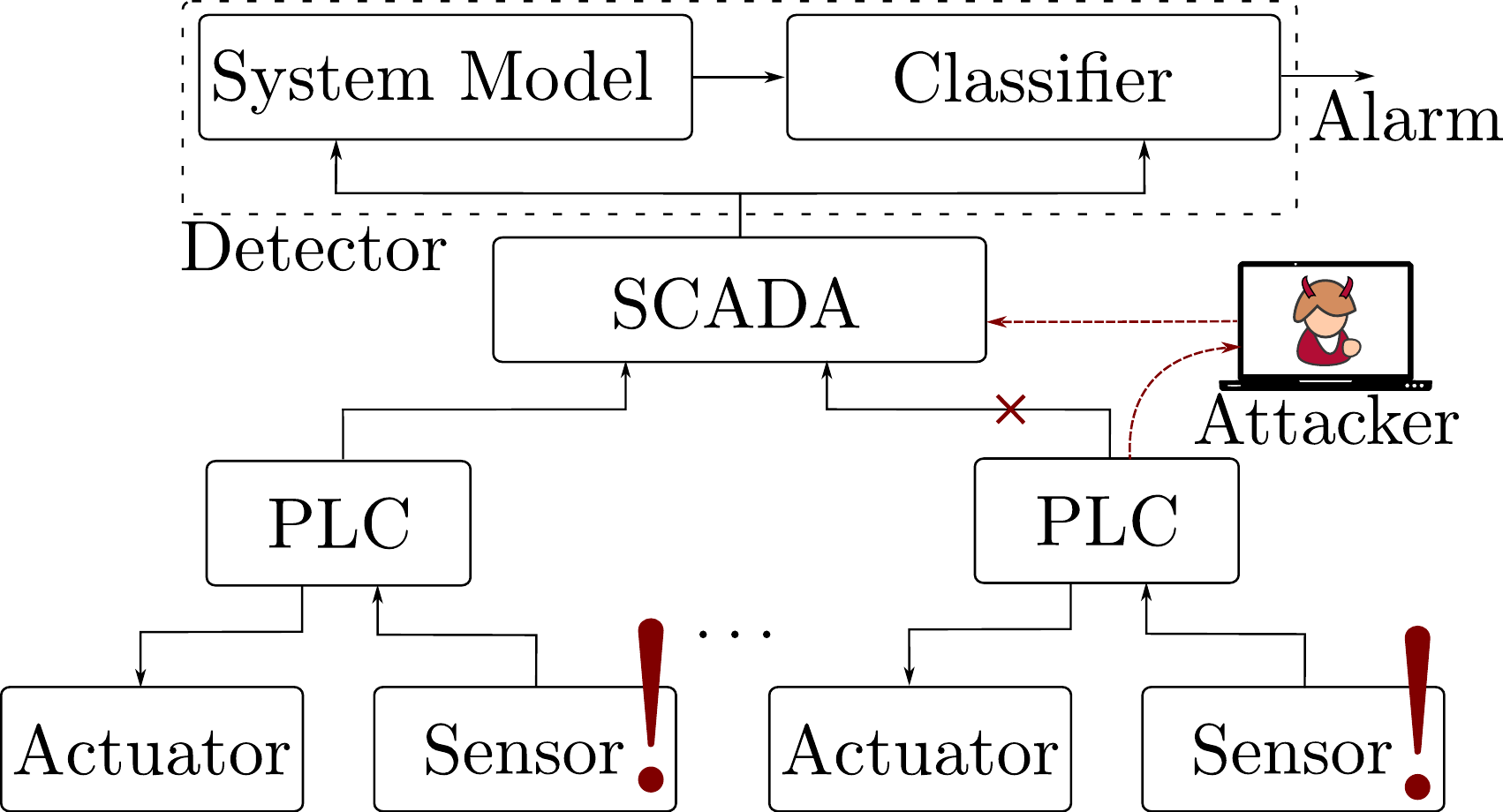}
    \caption{High level system and attacker model. The PLCs report sensor data about the anomalous process to the SCADA. The attacker can eavesdrop and manipulate a subset of the data provided to the SCADA. The reconstruction-based detector attempts to detect attacks based on a learned model of the system's benign operations.}
      %  \end{center}
\label{fig:attacker_model}
\end{figure}

\section{Concealment Attacks on Reconstruction-based Anomaly Detection} 
\label{sec:problem}
In this section, we introduce our system and attacker model, and our general problem statement for constrained concealment attacks. Then, we present our abstract white and \blackbox attacker model.

\subsection{System Model}
We consider a system under attack (Figure~\ref{fig:attacker_model}) consisting of several sensors and actuators connected to one or more PLCs, which are in turn connected to a SCADA system that gathers data from the PLCs. In our work, we assume that the SCADA is passive, so it does not send control commands to the PLCs (e.g., to actively probe for manipulations), or uses steganographic approaches to authenticate sensor readings~\cite{mo2015physical}. The SCADA feeds an attack detection system, whose goal is to accurately identify the instances in which the attacker manipulates the physical process while minimizing the number of false detections. The attack detection system generally consists of two main components: a \emph{system model}, which is used to generate additional features, and a \emph{classifier}, which (for each time step) classifies the system as either under attack or under normal operating conditions (see  Section~\ref{sec:related} for more details on classifiers in this context). During the attack, the physical process may be in an anomalous state, which will be detected unless the attacker manages to conceal it. The anomalies themselves are out of the scope of this work; we use prior-work datasets~\cite{taormina18battle, wadi17dataset}. 

To the best of our knowledge, our work is the first one that enables the use of constraints on the number of sensors that can be manipulated by an attacker (see Section~\ref{sec:attacker_model}).
As we will show, the performance of the attack degrades when lowering the number of channels that are under the attacker's control. Fully authenticated sensor signals would eventually prevent the attack to occur at the process level, but would impose cost on the normal system operations. Since our attacks exploit sensor signals received by the detector, they can be deployed somewhere else w.r.t. the industrial plant. Software exploits on the machine running the detector or historian server could also offer the attack surface to mount our proposed concealment attacks (those alternative attacker models are not modeled here for the sake of simplicity).

\subsection{Attacker Model}
\label{sec:attacker_model}

\subsubsection{Attacker Goal and Capabilities}
The goal of the attacker is to launch a concealment attack on an ICS to hide the real state of the process from an anomaly detector. The modeled attacker is assumed to have access to the ICS network, e.g., by physically attaching malicious devices to the network, intercepting communications to selected remote substations, or performing a Man-in-the-PLC~\cite{garcia17hey} attack. The attacker is thus assumed to control a subset of the communication between PLCs and the SCADA s\textbf{}ystem, and as a result, able to eavesdrop on traffic and send manipulated sensor readings to the detector. PLC communication with the SCADA system can be exploited to hide the real state of the system as practically demonstrated in~\cite{garcia17hey}. In contrast to~\cite{garcia17hey}, our attacker does not require explicit knowledge of the physical model equations to conceal the anomalies.

In particular, we assume that the anomalous physical process results in a feature vector $\vec{x}$, which triggers the detection system. The attacker thus needs to find an alternative vector $\vec{x}'$, which prevents detection of the attack. We formalize the \emph{concelament attack} as follows: given a feature vector $\vec{x}$ and a classification function $y()$ s.t. the detector correctly classifies $y(\vec{x}) =$  `under attack', the attacker is looking for a perturbation %$\vec{x}= [r_1, r_2,\dots, r_{n}]$ to 
$\vec{x}+\vec{\delta}$ s.t. $y(\vec{x}+\vec{\delta}) = $`safe'. Since the attacker's goal is to evade Mean Squared Error-based classifiers, the adversarial  attack has to find a perturbation $\vec{\delta}$ to minimize the reconstruction error between the input $\vec{x}+\delta$ and output $\hat{\vec{x}}$ of the Reconstruction-based detector. Please refer to Section~\ref{sec:attackedanomalydetector} for further details on the target model. In a mathematical notation, it can be written as the following constrained optimization problem in Equation~\ref{eq:opt_prob}:

\begin{equation}
\begin{aligned}
\label{eq:opt_prob}
	\text{minimize} \quad &  MSE={\frac{1}{n}\sum_{i=1}^{n}(\hat{x_i} - (x_i+\delta_i))^{2}}\\
	\textrm{s.t.}  \quad & \vec{\delta} \in \text{constraint space (Section~\ref{subsub:attacker_knowledge}) }\\
	& \text{real-time constraints imposed by CPS}\\
	& y(\vec{x}+\vec{\delta}) = \text{`safe'}
\end{aligned}
\end{equation}

We note that the attacks we demonstrate are not necessarily optimal, as the constraints  are satisfied with non unique solutions.
The attacks are conducted in real-time (i.e., in milliseconds per time step), not \emph{a posteriori} (i.e., applied retrospectively to a longer sequence of sensor readings after the attacker fully receives them).

\subsubsection{Attacker Knowledge}
\label{subsub:attacker_knowledge}
Using the adversarial learning notation introduced in  Section~\ref{sec:background}, a concealment attack is characterized by the knowledge of the attacker about the training dataset $\mathcal{D}$, feature set $\mathcal{X}$, learning algorithm $f$, and trained parameters $w$.
In the ICS setting, the attacker can be characterized differently according to his knowledge of the attacked system. In order to explain our attacker model, we split the tuple $(\mathcal{D}, \mathcal{X}, f, w)$ into two: the Data tuple $(\mathcal{D}, \mathcal{X})$ and Defense tuple $(f, w)$. We assume the attacker to be \emph{unconstrained} or \emph{constrained} w.r.t. the Data tuple, i.e.,~the sensor readings $\mathcal{X}$ that she can observe and manipulate and the data $\mathcal{D}$ that she eavesdrops. Moreover, we classify attackers as \emph{\whitebox}, \emph{\blackbox}, w.r.t. the Defense tuple, i.e.,~the knowledge of learning algorithm $f$, and trained parameters $w$. Table~\ref{tab:taxonomy}, provides an overview of the attacker's constraints considered in this work.

\begin{table}[t]
\caption{Classification of our attacker models based on training data and features, Data tuple $(\mathcal{D},\mathcal{X})$ and algorithm knowledge and parameters.  Access:\protect\includegraphics[width=0.4cm]{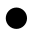}=full, \protect\includegraphics[width=0.4cm]{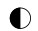}=partial. For all \whitebox attacks, the Defense tuple is $(f,w)$, for all \blackbox attacks, (\raisebox{-0.2cm}{\protect\includegraphics[width=0.4cm]{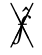}}, \raisebox{-0.1cm}{\protect\includegraphics[width=0.6cm]{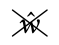}}).}
\setlength{\tabcolsep}{4pt}
\renewcommand{\arraystretch}{.75}
    \centering
    \begin{tabular}{rcccc}
        \toprule
          Attacker's&&   &  \multicolumn{2}{c}{$\mathcal{X}$}  \\
         \cmidrule(l){4-5} 
         Constraints & & $\mathcal{D}$ & Read& Write  \\  \midrule
          Unconstrained &\S~\ref{sec:concealmentresults}  & \begin{imageonly}\tDot\end{imageonly} & \begin{imageonly}\tDot\end{imageonly} & \begin{imageonly}\tDot\end{imageonly}  \\
         $\mathcal{X}$\xspace Partially &\S~\ref{sec:constrainded}   & \begin{imageonly}\tDot\end{imageonly} & \begin{imageonly}\tDot\end{imageonly} & \begin{imageonly}\thdot\end{imageonly} \\
        $\mathcal{X}$\xspace Fully& \S~\ref{sec:constrainded}           & \begin{imageonly}\tDot\end{imageonly} & \begin{imageonly}\thdot\end{imageonly} & \begin{imageonly}\thdot\end{imageonly} \\
       $\mathcal{D}$ &\S~\ref{sec:constrainded}           & \begin{imageonly}\thdot\end{imageonly} & \begin{imageonly}\tDot\end{imageonly} & \begin{imageonly}\tDot\end{imageonly} \\
       \bottomrule
      
    \end{tabular}
\label{tab:taxonomy}
\end{table}

\Paragraph{Constraints over Data Tuple} According to the Data tuple $(\mathcal{D}, \mathcal{X})$, we classify the attacker as:
\begin{itemize}
    \item\emph{Unconstrained} $(\mathcal{D}, \mathcal{X})$, in which the attacker can manipulate all $n$ features in $\vec{x}$, and her perturbations are limited in terms of \emph{L\textsubscript{0}} distance to be at most $n$. 

    \item\emph{Features Partially Constrained} $(\mathcal{D}, \mathcal{\hat{X}})$, we assume that the attacker is constrained to perturb a subset of $k$ out of $n$ variables in $\vec{x}$, and her perturbations are limited in terms of \emph{L\textsubscript{0}} distance to not exceed distance $k$. 
    
    \item\emph{Features Fully-Constrained} $(\mathcal{D}, \mathcal{\hat{X}})$, we assume that the attacker is constrained to observe and perturb a subset of $k$ out of $n$ variables in $\vec{x}$, and her perturbations are limited in terms of \emph{L\textsubscript{0}} distance to not exceed distance $k$. 
    
    \item\emph{Data Constrained} $(\mathcal{\hat{D}}, \mathcal{X})$, we assume that the attacker is constrained to eavesdrop a limited quantity of process data that are used for training its attacks. 
\end{itemize}

\emph{Selection of Constrained Features} The subset of features that can be modified is highly use-case dependent (i.e., which link is attacked, which device was compromised). To demonstrate the generality of our findings, we explored two types of constraints: a best-case scenario and a topology-based scenario.

For the best-case scenario, we assume that the selection of the $k$ out of $n$ manipulated features can be made by the attacker to maximize the attack impact. This arguably represents a best-case scenario for the constrained attacker (i.e., an attacker constrained to features that happen to be relatively ideal for the attacker). 
For the second scenario, constraints are derived from the network topology. We assume that the attacker can compromise a single substation (or PLC) in the network, and the selection of $k$ out of $n$ features is based on which sensors are interconnected to the compromised substation.

\Paragraph{Knowledge of Defense Tuple} We classify the attacker according to their knowledge of the Defense tuple $(f,w)$, as:
\begin{itemize}
    \item\emph{White box} $(f, w)$, the attacker knows the exact system model and its variables (such as the currently estimated system state), and the exact thresholds of the classification system.  Thus, the \whitebox attacker is characterized by the knowledge of $(f, w)$. With that information, the attacker could either run a basic exhaustive search, basic optimization strategies, or more sophisticated approaches (especially solutions that use the gradient signal from the attacked model).

    \item\emph{Black box} (\raisebox{-0.2cm}{\protect\includegraphics[width=0.4cm]{image3.pdf}}, \raisebox{-0.1cm}{\protect\includegraphics[width=0.6cm]{image4.pdf}}), the attacker is aware of the general detection scheme, but unaware of internal variables, architecture and exact thresholds used in the classification. We note that our \blackbox attacker is different from the one defined in~\cite{biggio18wild}, $(\hat{f}, \hat{w})$. Our attacker does not require the knowledge of $f$ or its approximation $\hat{f}$. In our case, the nature of the environment imposes that the attacker cannot query the system even in a \blackbox manner to get feedback on the provided labels or confidence scores (this is done for example in~\cite{tramer2016stealing, xu2016automatically, chen2017zoo, dang2017evading}), as this would mean potentially raising the alarm. Thus, we consider that the only assumption of the attacker concerning $f$ is that Deep Learning techniques are used for detection. 
\end{itemize}

Given this taxonomy, the attacker can be classified
for example, as \emph{unconstrained \whitebox}.

\subsection{Example Constraint Scenarios}

We argue our \emph{Constrained} and \emph{Unconstrained} attacks represent a realistic threat model in the ICS setting and fit the taxonomy of attacks in the AML literature. In particular, practical ICS are typically composed of multiple stages, and each stage is controlled by a different PLC (i.e., different brands/models). Moreover, the ICS can be deployed in a physically distributed manner. For example, in the case of water distribution networks, pumping stations are typically located kilometers away from the water reservoir. In this heterogeneous setting, an attacker can either gain control over a limited set of resources as practically demonstrated in~\cite{garcia17hey}, or the whole plant (by compromising the network or central SCADA). We capture those different capabilities of the attacker in the following three scenarios:
\begin{itemize}
    \item The \textit{Unconstrained Attacker} can read and write any features arbitrarily, e.g., by compromising the SCADA system, as modeled by Mo et al.~\cite{mo2009secure}. 
    \item The \textit{Partially Constrained attacker} can read all traffic received by the SCADA system (for example by a passive wiretap~\cite{wiretap} or by leveraging access control misconfigurations), but she is only able to spoof sensor readings from a specific substation, exploiting specific vulnerabilities of the substation or its protocol to the SCADA (e.g., lack of authentication). 
    \item The \textit{Fully Constrained attacker} relates to a scenario where an attacker compromised a specific substation, giving him read and write access to features from this substation only~\cite{abrams2008malicious, garcia17hey}.
\end{itemize}

Similar assumptions were also considered in the BATADAL dataset~\cite{taormina18battle}, where the attacker was assumed to perform constrained replay attacks to reduce the confidence of anomaly detectors. In our contribution, we perform a concealment attack in a systematic way to assess their impact over the anomaly detector.
 
A similar intuition is also used in the FAIL attacker model~\cite{suciu2018does} for AML. In particular, FAIL proposes to characterize attacker knowledge over 4 dimension: \textbf{F}eature, \textbf{A}lgorithm, \textbf{I}nstance, \textbf{L}everage. While the first three have a counterpart in systematization by Biggio et al.~\cite{biggio18wild}, \textbf{L}everage stands for the subset of features that the adversary can modify (just like our constrained attacker). Thus, our \emph{Unconstrained} and \emph{Constrained} attacks represent attacks with full and limited \textbf{L}everage.

\subsection{Our Framework for Attack Computation}
For both the \whitebox and \blackbox case, the attacker is assumed to intercept and manipulate a Constrained or Unconstrained set of sensor readings in real-time.

In the \whitebox setting, we propose an \approachI attack, able to interactively query a classification oracle to determine which features to manipulate, and the value to assign to those features. We propose to compute the manipulations using an iterative algorithm. This algorithm calculates solutions that are `safe' from the detector perspective. The algorithm is tunable, i.e., the attacker can act on some algorithm parameters that impact over time the computation and, consequently, the concealment efficacy. Again, this speeds up computation but can impact the solution quality.

In the \blackbox setting, we propose the use of a \approachII attack, specifically a Deep Neural Network that is capable of outputting concealed sensor readings, without the oracle's feedback. The attacker is adversarially training the neural network to learn how the detector expects the ICS to behave. This trained neural network then receives the traffic coming from the PLC. While the attacker creates an anomaly over the physical process, the neural network adjusts the anomalous data to resemble `safe' data. This manipulated version of sensor data is sent to the SCADA. 

In order to avoid confusion, we point out explicitly that our \approachI attack can operate under the attacker \whitebox assumption as it requires to query an oracle of the anomaly detector, while the \approachII attack can operate under the \blackbox assumption as no query access is required to compute the adversarial sensor readings.
We compare \approachI and \approachII approaches with replay attacks, as proposed in literature~\cite{mo2009secure}. The attacker that performs a replay attack can be categorized as \blackbox.

\section{Design of Concealment Attacks}
\label{sec:design}
We now present a detailed design for the three attacks that we consider. We start with details on the (prior work) Reconstruction-based attack detector, then introduce the replay attack (proposed by~\cite{mo2009secure}). We provide details on the \approachI attack (\whitebox knowledge). We then conclude with the \approachII approach (\blackbox knowledge), which leverages an online concealment method without any prior knowledge about the physical process that generates the sensor readings and the detection scheme. Given these premises, we note that, while adversarial examples found using the \approachI approach depend on the internal structure of the attacked anomaly detector, examples crafted through the \approachII approach are independent from the addressed detection scheme (see Section~\ref{sec:generalize}).

\subsection{Background: Reconstruction-based Attack Detector}
\label{sec:attackedanomalydetector}
In this work, we target anomaly detection systems proposed in prior works~\cite{goh2017anomaly, kravchik2018detecting, taormina2018deep}, which share the same underlying idea, reconstruction-based anomaly detection. 
 
The anomaly detector consists of two parts, namely a Deep Learning autoencoder model (with $m\times n$ features as input and $n$ output) trained over the normal operation sensor readings of an ICS to optimize Mean Squared Error Loss, and a classifier function. The idea is that the deep model has learned to reproduce the system behavior under normal operating conditions with a low reconstruction error, so it reproduces a higher reconstruction error when fed with anomalous sensor readings. The comparison between the input and output of the deep model is used to decide if the system is `safe' or `under attack'. Reconstruction based classifiers represent the state of the art for anomaly detection in ICS on a multi-fold basis. First, they can overcome the problem of shortage of `under attack' samples that are hard to be gathered from the system without damaging the plant. Second, they can capture interdependence between sensor signals that helps localization of anomalies. Third, they guarantee a low time of detection, which is a fundamental property for ICS anomaly detectors.

As reference implementation, we use the general Autoencoder-based anomaly detector framework proposed in~\cite{taormina2018deep} and available as open source~\cite{aeed18repository}. Moreover, we explore transferability of our \blackbox attack between different Deep Architectures (DA) in Section~\ref{sec:implementation}. In particular we tested Long Short Term Memory (LSTM)~\cite{hochreiter1997long} as proposed in~\cite{goh2017anomaly} and Convolutional Neural Networks (CNN)~\cite{krizhevsky2012imagenet} as proposed in~\cite{kravchik2018detecting}. The input to the  DA is $X=[\vec{s_{t-m}}, \dots,  \vec{s_t}]$, representing $m+1$ time-steps of $\vec{s} = [r_1, r_2, \dots, r_n]$, which is an $n$-dimensional vector of sensor readings. DA's goal is to find $\phi$ parameters that minimize following the Mean-Squared Error optimization problem:
\begin{equation}
\label{eq:training}
    \min_\phi
    \mathbb{E}_{(X,\vec{s_t})\sim\mathcal{D}}\left[
    \lVert DA(\phi,X),\vec{s_t}\rVert_2^2\right]\; 
\end{equation}

where \emph{DA} outputs an n-dimensional vector $\vec{o} = [v_1, v_2, ..., v_{n}]$, and $v_i \; s.t. \; i \in \{1,...,n\}$ represents the reconstructed value w.r.t. the input reading $r^t_i$ . In order to decide if the system is under attack, the mean squared reconstruction error between observed and predicted features are computed. If the mean squared reconstruction error exceeds a threshold $\theta$, the system is classified as under attack. The authors of~\cite{taormina2018deep} chose $\theta$ as  99.5 percentile (Q99.5) of the average reconstruction error over the training set. 

We formalize this as follows. Given a target vector $\vec{s_t}$, we define: $\vec{e}=\vec{s_t}-\vec{o}=[d_1, \dots, d_n]$ as the reconstruction error $n$-dimensional vector,
$\varepsilon(\vec{e})$ as the corresponding average reconstruction error:
\begin{equation}
\varepsilon(\vec{e}) = \frac{1}{n}\sum_{i = 1}^{n}  {d_i}^{2},
\end{equation}
and $y(X)$ as the classified state of the water distribution system out of Reconstruction-based Intrusion Detection System. Given an input $X$, $y$ is `under attack' if $\varepsilon$ greater than $\theta$:
\begin{equation} y(X)=
\begin{cases}
\begin{aligned}
    \text{`under attack' if} \;\varepsilon(\vec{e})  > \theta \\[1ex]
  \text{`safe' otherwise}
\end{aligned}
\end{cases}
\end{equation}
Moreover, the authors propose a \emph{window} parameter that takes into consideration the mean of $\varepsilon(\vec{e})$ of the last \emph{window} time steps to decide if the current tuple is `safe'. This helps diminishing the amount of false positives, since an alarm is raised only if in the last \emph{window} time steps the mean of $\varepsilon(\vec{e})$ is above $\theta$.

\subsection{Baseline: Replay Attack}

In the replay attack setting (prior work,  used here as baseline), the attacker does not know how detection is performed. In order to avoid detection, the attacker can replay sensor readings that have been recorded while no anomalies were occurring in the system. In particular, we assume that the attacker could record selected data occurring exactly $n$ days before---i.e., if the concealment attack starts at 10 a.m., the attacker starts replaying data from 10 a.m. one day before.
 
\subsection{Iterative Attack}
\label{sec:withebox}
In the \approachI attack, the \whitebox attacker knows how detection is performed, all thresholds and parameters of the detector, as well as the normal operation range for each one of the model features. For example, the attacker knows which sensor readings are common during normal operation of the physical process. As a result, the attacker essentially has access to an \emph{oracle} of the Deep Architecture, where the attacker can provide arbitrary $\vec{x}$ features and gets the individual values of the reconstruction error vector $\vec{e}$. The attacker then computes $\max_i\vec{e}$ and finds the sensor reading $r_i$ with the highest reconstruction error from $\vec{x}$. 

In order to satisfy the constraint $\varepsilon(\vec{e'}) < \theta$ (i.e. $y(\vec{x}+\vec{\delta}) = \text{`safe'}$ in Equation~\ref{eq:opt_prob}), the attacker  performs a coordinate descent algorithm to decrease the reconstruction error related to $r_i$ (As we rely on coordinate descent algorithm we do not use gradient estimation methods). At each iteration of the algorithm, a coordinate of the feature vector is modified until a solution is found or the computational budgets are exceeded. 

Two computational budgets are put in place: \emph{patience} and \emph{budget}. If no lower reconstruction error is found by descending a coordinate, the algorithm tries descending other coordinates. If no improved solutions are found in \emph{patience} iterations, the input is no longer optimized. \emph{budget} is the maximum number of iterations for coordinate descent.  After \emph{budget} attempts without satisfying $\varepsilon(\vec{e}') < \theta$, the input is no longer optimized, and no solution is found. Additional details  are found in Appendix~\ref{sec:iterativeattackdetails}.

Sensor readings $r_i$ are modified in the range of normal operating values; this guides the computation to a solution that is consistent with the physical process learned by the detector. For example, if normal operations of sensor $r_i$ are in the range $[0, 5]$, the attacker tries to substitute the corresponding value of $r_i$ according to its range to see if the related reconstruction error decreases. The algorithm~\ref{alg:WEM} can be found in Appendix~\ref{sec:iterativeattackdetails} as a reference.

\subsection{Learning Based attack}

In the \approachII setting, the \blackbox attacker does not know anything about the detection mechanism except the fact that it relies on a Reconstruction-based model: the attacker can only intercept and manipulate the communication between the PLCs and SCADA. However, the nature of the ICS environment allows us to assume that a detection mechanism trained over a specific CPS should represent its physical rules in order to spot anomalies.

In this case, a reasonable attack scheme can be divided into five steps. The attacker first intercepts traffic from PLCs to SCADA in order to collect information on how the ICS behaves under normal conditions. Second, collected data are used to learn how the system behaves normally and train a Deep Learning model, implicitly training a model to solve the minimization problem in Equation~\ref{eq:opt_prob}. Third, the attacker manipulates the physical process; anomalous data are generated as a consequence. Fourth, the adversarially trained model is used to conceal anomalous readings by morphing them into concealed data that will be classified as `safe'. Fifth, the concealed data is forwarded to the SCADA.

\Paragraph{Autoencoder-based Generator} We implement the \approachII attack using an autoencoder network to generate concealed data (the word generator is used with a different meaning than the usual one. In our case the input is not random noise that is going to be crafted by the network). The autoencoder is trained while intercepting normal traffic; the network learns to output tuples that are classified as being normal with high confidence. Forwarding the output of the adversarial network---regardless of how the detector is built---forces it to misprediction, because the adversarial examples have been adjusted to resemble normal operations. Note that the autoencoder does not completely change the tuple; intuitively, only the part that contains the anomaly is reconstructed to match the learned physical behavior.
We explored different architectures, activation, and loss functions for the autoencoder network. Experimentally, we verified that by using hidden layers to increase the data dimensionality (i.e.,~with more units than the input and output layers), rather than decreasing it, we achieved a higher concealment capacity than canonical `compressing' models. Particularly, we implemented an autoencoder network with three hidden layers, with input and output dimensions equal to the number of sensors and actuators in the network. We used mean squared error as a loss function and sigmoid as an activation function. To train the network, we use the ADAM~\cite{kingma2014adam} optimizer with a learning rate set to $0.001$.

\Paragraph{Post-processing} In order to generate feasible inputs for the anomaly detector, we need to consider that not all the sensor readings assume continuous values---some are categorical integers that represent the status of actuators. Since the output of a neural network is continuous, we need to post-process all the readings that are supposed to be integers. For example, if a pump status assumes a value $0$ when it is turned \emph{off} and $1$ when it is turned \emph{on}, post-processing approximates the corresponding output value to the nearest allowed integer. According to this post-processing, some other values should be adjusted in order to match the physical rules. This is the case, for example, for speed sensors that must read $0$ if their related pump is \emph{off}.

\subsection{Positioning with respect to State-of-The-Art AML attacks}\label{sec:positioning}
In this work, we consider attacks on a (prior work) reconstruction based classifier. This represents a substantial difference from classifiers considered in AML attacks in other domains. In related work~\cite{papernot16eurosp, carlini17sp, MadryMSTV18}, attacks on end-to-end Neural Network classifiers are considered (i.e., those classifiers are trained to optimize the cross-entropy loss). In particular, target misclassification is achieved,  diminishing the predicted probability of the true class in the output layer. Our problem setting is different: To evade the classifier, we need to diminish the residual between input and output. Our target Neural Networks are trained to optimize the Mean Squared Error loss. This can be achieved by reconstructing the sensor signal in a way that matches the learned physical properties of the ICS. Those differences motivated our novel white-box and \blackbox approaches to evade Deep Learning-based Anomaly Detectors in ICS.

\Paragraph{Model Robustness}
Adversarial Robustness~\cite{MadryMSTV18} is achieved by a) adversarial training b) classifier capacity. Following the definition, adversarial training is obtained embedding adversarial examples in the training set. In our context, the Neural Network is trained to approximate the system behavior during normal operating conditions, with no samples for the `under attack' class. Thus, adversarial training here does not apply: indeed, we cannot train the system to be resilient to adversarial attacks since samples from the class `under attack' is unknown to the defender.

\section{Evaluation}
\label{sec:implementation}
In this section, we experimentally evaluate the proposed attacks. We start by introducing the datasets we used for our experiments: the BATADAL dataset and data coming from a real industrial process (WADI dataset). We start the evaluation targeting an Autoencoder-based detector and explore the performance of replay, \approachI{}, and \approachII{} attacks in constrained and unconstrained conditions. Then, we show that our \approachII{} attack generalizes to other schemes based on LSTM and 
CNN. Finally, we show the concealment attack results obtained in a real industrial testbed.

\subsection{Dataset 1: BATADAL} 
\label{sec:epanetCPAandBATADAL}
The first dataset was generated with epanetCPA~\cite{taormina2019epanetCPA}, an open-source, object-oriented Matlab toolbox for modeling the hydraulic response of water distribution systems to cyber-physical attacks. The dataset was originally generated for the BATADAL~\cite{taormina18battle} competition, which ran between 2016  and 2017. 
The BATADAL competition was based on three datasets: the first contains data coming from the simulation of 365 days of normal operations, while the second and third contains 14 attacks (7 attacks each). The details of the attacks can be found in~\cite{taormina18battle}. 
These datasets contain readings from 43 sensors: tank water levels (7 variables), inlet and outlet pressure for one actuated valve and all pumping stations (12 variables), as well as their flow and status (24 variables).
All variables are continuous, except for the status of valve and pumps, represented by binary variables. 

The original attack dataset (from \url{http://www.batadal.net/data.html})  contained sensor data readings that were manually concealed. For that reason, we could not use the original attack dataset directly (as we wanted to add concealment ourselves). Instead, we re-created the attacks (and resulting sensor data) from the BATADAL dataset for this work using the original setup, without any manual concealment. In our new version, the data are collected from sensors every 15 minutes. 

\subsection{Dataset 2: WADI}
Our second dataset is based on the Water Distribution (WADI) testbed, a real-world ICS testbed located at the Singapore University of Technology and Design~\cite{chuadhry17wadi}. It is composed of two elevated reservoir tanks, six consumer tanks, two raw water tanks, and a return tank. It contains chemical dosing systems, booster pumps and valves, instrumentation, and analyzers. WADI is controlled by 3 PLCs that operate over 103 network sensors.
Moreover, the testbed is equipped with a SCADA system. WADI consists of three main processes: P1 (Primary supply and analysis), P2 (Elevated reservoir with Domestic grid and leak detection), and P3 (Return process). For anomaly detection purposes, we consider sensor data from P1 and P2, since the return process is only implemented for recycling water. Considering stages P1 and P2, we have data coming every second from 82 sensors. 
In this work, we use two WADI datasets. The first dataset contains data of 14 days of normal operations. The second contains 15 attacks on physical processes spanned over two days of operations. This dataset is available on request~\cite{wadi17dataset}.

We primarily use the WADI dataset for two reasons: i) to show that the discussed detection mechanism applies to real-world ICS data, and ii) to see whether our attack methodology is transferable from a scenario in which simulated data are used to another scenario in which real data are used. %We summarize the results of our attacks on the WADI scenario next.

\subsection{Evaluation Setup}

We evaluate the detection Recall over datasets under original conditions (i.e., no concealment attacks), replay, iterative, and learning-based concealment attacks. The $Recall$ (True Positive Rate) is defined as: 
\begin{equation}
Recall =
   \frac{TP}{TP+FN},
\end{equation}
where $TP$ stands for True Positive and $FN$ for False Negative.
Recall measures the rate of correctly classified positive instances. When the anomaly detector is tuned, a higher Recall means that the anomaly detector is correctly retrieving anomalies. The attacker's concealment goal can be expressed in terms of $Recall$: the concealment attack is successful if the detector Recall over the concealed tuples goes to $0$. The closer the Recall comes to $0$, the higher the amount of misclassified tuples. Note that we launch our concealment attacks over the instances of anomalous data, i.e.,~data reporting ground truth `under attack'.

Both \approachI and \approachII attacks are implemented using Python 3.7.1; neural networks are implemented and trained using Keras 2.3.1 with TensorFlow 1.11.0 backend. Experiments use a laptop equipped with Intel i7-7500U CPU, 16GB of RAM, and NVIDIA GeForce 940MX GPU 4GB.

\Paragraph{Training of Attack Detector} For both BATADAL and WADI, we trained the third party attack detector~\cite{taormina2018deep} on sensor readings occurring during normal operational data. 

For the BATADAL dataset (where sensor readings are sampled every 15 minutes), we found that parameter \emph{window}~=~$3$ quarter of hours is a reasonable decision boundary to flag correctly attacks and do not raise False alarms. This gives a \emph{Accuracy}~=~$0.93$, \emph{Precision}~=~$0.90$, \emph{Recall}~=~$0.60$, \emph{FPR}~=~$0.01$. Changing the $window$ parameter, we can increase the \emph{Recall} at the price of decreasing \emph{Precision} that means raising a higher number of False Alarms.

For the WADI dataset, we found that parameter \emph{window}~=~$60$ seconds is a reasonable decision boundary to flag correctly attacks and distinguish them from False Positives. This gives a \emph{Accuracy}~=~$0.97$, \emph{Precision}~=~$0.77$, \emph{Recall}~=~$0.68$, \emph{FPR}~=~$0.01$. 

Results are in line with the current state of the art detection over the BATADAL dataset~\cite{taormina2018deep} and WADI~\cite{feng2019systematic}.

\Paragraph{Replay attack} In this attack, the attacker replays for the whole duration of the physical manipulation, using the sensor readings as recorded at the same hour $s$ days before (assuming that process operations are often periodic within 24h). $s$ is chosen to let the replay contain only normal operations data. For example, given a physical manipulation that lasts 50 hours, we replay sensor readings as happened 72 hours earlier.

\Paragraph{Iterative attack} The attacker manipulates variables required to find a solution (according to the two stopping criteria introduced in Section~\ref{sec:withebox} and constraints over modifiable sensor readings). For BATADAL dataset, we tuned the two stopping criteria via grid search to guarantee a trade-off between the decrease of detection accuracy and computational time. Specifically we selected \emph{patience}~=~$15$ and the \emph{budget}~=~$200$. For WADI dataset, the \approachI parameters (following the same rational as in BATADAL case) we choose are $\text{\emph{patience}} = 40$ and the $\text{\emph{budget}} = 300$. The result of this experiment depends on the detection mechanism. The attacker is using the oracle to determine if the concealment is successful. 

\Paragraph{Learning based attack} For the \approachII attack, the attacker uses an autoencoder (\emph{AE}) as the generator and sends predicted readings to the SCADA. According to the attacker's constraints, we train an autoencoder over the readable features. We used sigmoid as activation function, Gorlot initialization ~\cite{glorot2010understanding} as weights initializer and mean squared error as loss function. Moreover, we split the data in train $\frac{2}{3}$ and validation $\frac{1}{3}$, use early stopping~\cite{earlystopping} to avoid overfitting and reduce learning rate on plateaus~\cite{ReduceLROnPlateau}.
Depending on the constrained scenario (i.e., the features that the attacker can read $\mathcal{X}$ or the amount of data that she spoofed $\mathcal{\hat{D}}$), the adversarially trained autoencoder has a different number of input neurons. Given $n$ as input/output dimension, the autoencoder is composed of 3 hidden-layers with respectively $2n$, $4n$, $2n$ neurons. To perform and evaluate the learning based attacks, we trained $83$ models with BATADAL data and $63$ for WADI. In the unconstrained case, for the BATADAL dataset ($43$ variables), we train an autoencoder with $64$ and $128$ units for the first/third and second hidden layers, respectively, training requires 18 epochs (2 seconds/epoch), for a total of 36 seconds to train the model; for the WADI dataset ($82$ variables), we use $128$ and $256$ units, training required 7 epochs (64 seconds/epoch), for a total of 488 seconds to train the model.  

\subsection{Unconstrained Concealment Attack}
\label{sec:concealmentresults}
\begin{table}
    \small
    \centering
    \caption{Detector Recall (BATADAL (B) and WADI (W) datasets), before and after unconstrained concealment attacks.
        The column `Original' refers to the detection Recall over the data without concealment; `Replay', reports the Recall after replay attack, while `Iterative' and `Learning based' columns report the Recall after our proposed adversarial concealment attacks.}
        \begin{tabular}{ccccc}
            \toprule
            & \multicolumn{4}{c}{Detection Recall} \\ 
            \cmidrule(l){2-5}
            Data & Original& Replay & Iterative &  Learning based \\\midrule
            B & ${0.60}$ & $ {0} $ & $ {0.14} $ & $ {0.14} $ \\
            W & ${0.68}$ &$ {0.07} $ & $ {0.07} $ & $ {0.31} $\\
            \bottomrule
        \end{tabular}
    \label{tab:BATADAL-WADI}
\end{table}
\begin{table}
    \small
    \centering
        \caption{Average required time (in seconds) to manipulate sensor readings. `Replay' column is empty as replay attacks do not require computation. 'Iterative' and `Learning based' columns report the mean and std deviation required to compute the manipulation sensor readings at a given time step.}
        \begin{tabular}{cccc|cc}
            \toprule
            &\multicolumn{5}{c}{Computational time, mean(\smean) and std(\sstd)}\\
            \cmidrule(l){2-6}
            & Replay &\multicolumn{2}{c}{Iterative} & \multicolumn{2}{c}{Learning based}  \\ 
            \cmidrule(l){3-4} \cmidrule(l){5-6} 
            Data &  &\smean[s] & \sstd & \smean[s] & \sstd
            \\ \midrule
            B & - & $ {2.28} $ & $ {2.46} $ & $ {0.002} $ & $ {0.005} $\\
            W & - & $ {0.60} $ & $ {0.41} $ & $ {0.005} $ & $ {0.002} $\\
            \bottomrule
        \end{tabular}
        \label{tab:BATADAL-WADI-time}
\end{table}

In this experiments, we assume the Unconstrained attacker $(\mathcal{D}, \mathcal{X})$ that is able to read and control all the reported sensor readings, in the White box $(f, w)$ and Black Box(\raisebox{-0.2cm}{\protect\includegraphics[width=0.4cm]{image3.pdf}}, \raisebox{-0.2cm}{\protect\includegraphics[width=0.4cm]{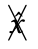}}) scenarios. We discuss the results of our evaluation of the detector for both datasets in several scenarios (see Table~\ref{tab:BATADAL-WADI}). We evaluated the performance of our concealment attacks over the time steps with ground truth `under attack' labels only, i.e., we exclude normal operation data time steps from the computation of Recall for this attack evaluation.

The first row of Table~\ref{tab:BATADAL-WADI} reports the average results obtained with the three different attack strategies. 
In this setting, the replay attack is giving $0$ Recall over the replayed sensor readings. This means that when the attacker can manipulate all the sensor readings, the anomaly detector is no more able to spot the attack occurring over the physical process. Considering the \approachI and \approachII approaches, we notice that the Recall is $0.14$, this represents a significant drop in detector performance, but not as effective as the replay of all sensors. 

The second row of Table~\ref{tab:BATADAL-WADI} refers to concealment attacks over the WADI dataset. 
The result over this dataset shows that the replay attack can hide the anomaly occurring over the CPS. The performance of the \approachI  equals the one of the replay attack. Finally, the \approachII approach is underperforming the other methods. Despite this, the detector's Recall reduces more than 50\% after \approachII manipulation.

\Paragraph{Computational Time} Table~\ref{tab:BATADAL-WADI-time} reports the average time required to compute the adversarial examples. In contrast to \approachI and \approachII, the replay attack does not require computation. The iterative approach requires an amount of time that depends on the algorithm computational budgets. The \blackbox approach requires a constant amount of time since it consists of a neural network prediction. Given our real-time constraints of adversarial examples computation (i.e., target time within milliseconds), we can conclude that \approachII approach easily meets the requirements. In the BATADAL case (where the sampling time is $15$ minutes), we do not require more than $2 ms$ on average to compute an adversarial example. In the WADI case (where sampling time is $1$ second), on average, we do not require more than $5 ms$ to compute an adversarial example. The iterative attack is slower, but on average, still below the sampling intervals.

\Paragraph{Summary of Unconstrained Attacks findings} When the attacker is free to manipulate all the sensor readings, results show that replay attacks hide anomalies occurring over the physical process. First, a replay attack does not require computation to find the manipulated set of sensor readings; second, the attacker does not need to be aware of the detection mechanism; and third, the considered anomaly detector Recall goes to zero since the replayed sensor readings do not contain (additional) anomalies. White box, even though it achieves valuable results, requires computation, and the attacker needs to be omniscient w.r.t. the defense mechanism. We note that the \approachII attack can decrease the detector's Recall without having access to detector's oracle, with low computational effort (after training) and the same knowledge w.r.t. the attacked model as the replay attack.

\subsection{Constrained Concealment Attack}
\label{sec:constrainded}
In the previous subsection, we found that full replay attacks can be a powerful and low-cost way to evade anomaly detectors if all features can be replayed. In this section, we demonstrate the impact of constraints on the attacker,  e.g., if the attacker can only control a subset of the reported sensor values. Specifically, we perform Partially, Fully, and Data constrained attacks as modeled in Section~\ref{sec:attacker_model} and show how our proposed \approachI and \approachII outperform replay attacks.

\begin{figure*}
    \centering
     \subfigure[BATADAL]{\includegraphics[scale=0.55]{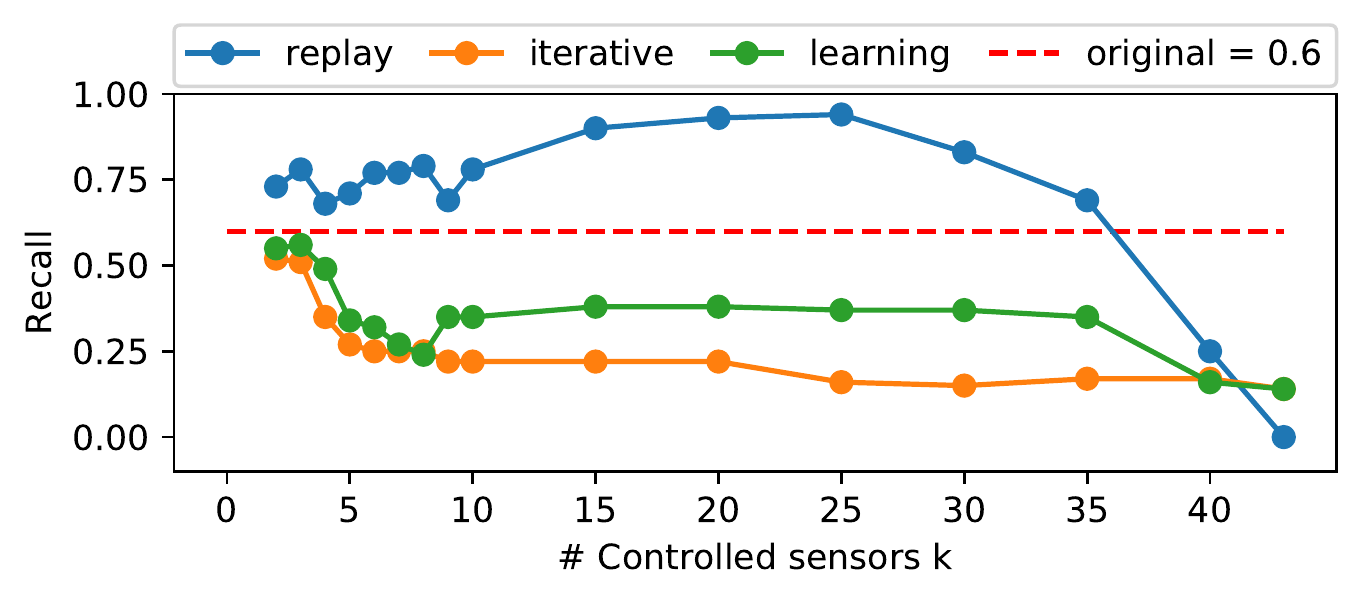}}
    \subfigure[WADI]{\includegraphics[scale=0.55]{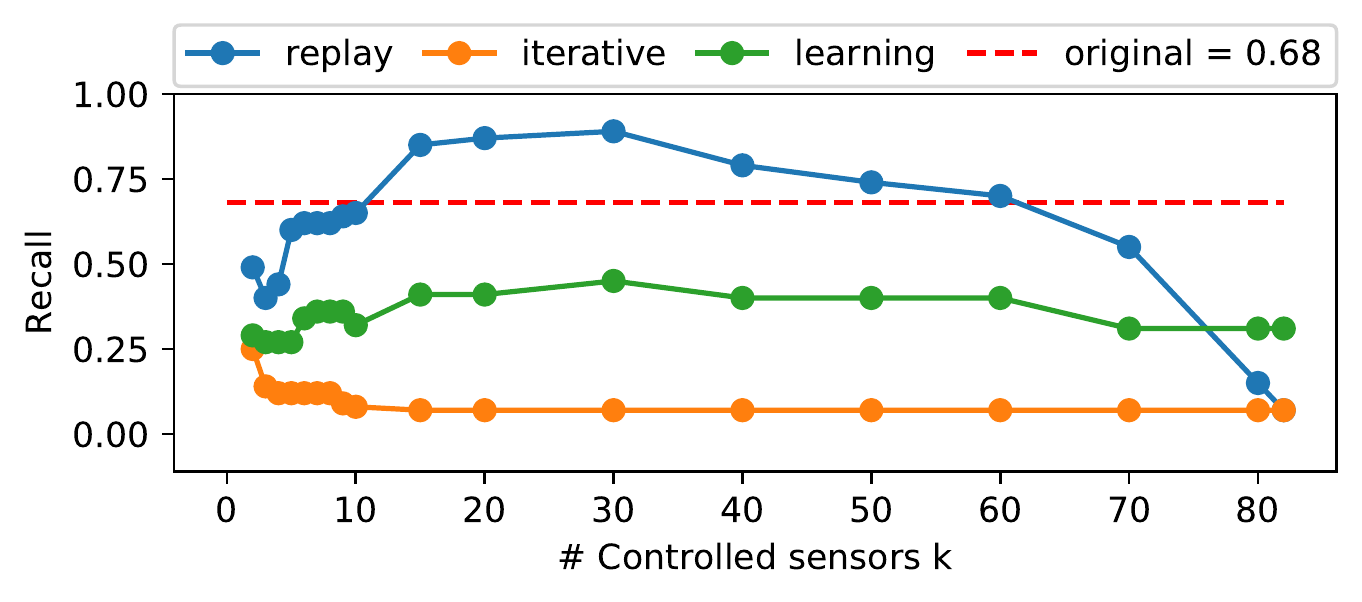}}
    \caption{Impact of Partially Constrained attacker (best-case scenario), comparison between replay attack and our proposed concealment attacks. In the constrained scenario, we notice that replay attack performs bad increasing detector's Recall and raising more alarms than the original data without concealment. This is due to contextual anomalies introduced as part of large-scale replay.  Both \approachII and \approachI approaches outperform the replay attack as they do not introduce contextual anomalies and reduce detector's Recall manipulating few features.}
    \label{fig:BATADAL-WADI-constrained_result}
\end{figure*}

\Paragraphs{Partially Feature-Constrained attack,} $(\mathcal{D}, \mathcal{\hat{X}})$.
Figure~\ref{fig:BATADAL-WADI-constrained_result} reports the average result of the constrained attacks over BATADAL and WADI datasets with an \emph{best-case} selection of constraints. Due to space limitations, the constraint selection can be found in Appendix~\ref{sec:constraintsdef}. In the case of the BATADAL dataset, we note that the replay attack does not cope well with constraints. Since the anomaly detector can spot the presence of contextual anomalies, the replay of only $k$ features results in alarms, with an average detection Recall higher than in the benign case (i.e., no replay of sensors applied), the value decreases when 40 out of 43 sensors are replayed. In the case of \approachI and \approachII attacks, we can notice that the detection Recall is always lower than the original Recall. In the \approachI case, Recall decreases with the number of features that can be modified. Learning-based attack Recall is not monotonically decreasing with the number of features that can be modified. Specific constraint sets  better match  the physical rules learned by the detector and allow the creation of more effective adversarial examples. In the case of WADI, we can observe that the replay attack can diminish the detector's Recall, especially when the attacker manipulates 3 or more features. The iterative based attack can achieve the same Recall as if in the Unconstrained Attack case when manipulating 15 out of 82 features. In the case of \approachII attack, results show that for 3 manipulated (best-case) features, the attack performs slightly better than in the unconstrained case.

In the case of \emph{topology-based} constraints, the attacker controls the sensors connected to 1 PLC in the network. We found that in the BATADAL case, Recall is reduced to $0.36$ with the replay attack and $0.34$ with the iterative and learning-based attack. In the WADI case, Recall increases to $0.64$ with the replay attack while it is reduced to $0.12$ in the iterative attack and $0.36$ in the learning-based attack. In addition, in this case the iterative and learning-based approach overcomes the limitations of constrained replay attacks, especially in the case of the WADI dataset (Table~\ref{tab:BATADAL-WADI-constrained_result} in the Appendix reports the numerical Recall scores found in the different constrained settings).

\Paragraphs{Fully Feature-Constrained Attacker,} $(\mathcal{D}, \mathcal{\hat{X}})$. In the case of the fully constrained attacker, Replay and Iterative attack approach do not change, since those two methods do not exploit correlations among features to output the perturbations. The learning-based attack is the only one affected by these constraints, i.e.,~the adversarial network is trained on the constrained set of sensor readings. We launched this attack with the topology based constraints, and we found that also in this case, when the attacker gains control of a PLC over the network, the detection Recall can be compromised by the attacker. In the BATADAL case, Recall is reduced to 0.39. In the WADI case is reduced to 0.45.

\begin{figure*}
    \centering
    \subfigure[LSTM]{\includegraphics[scale=0.55]{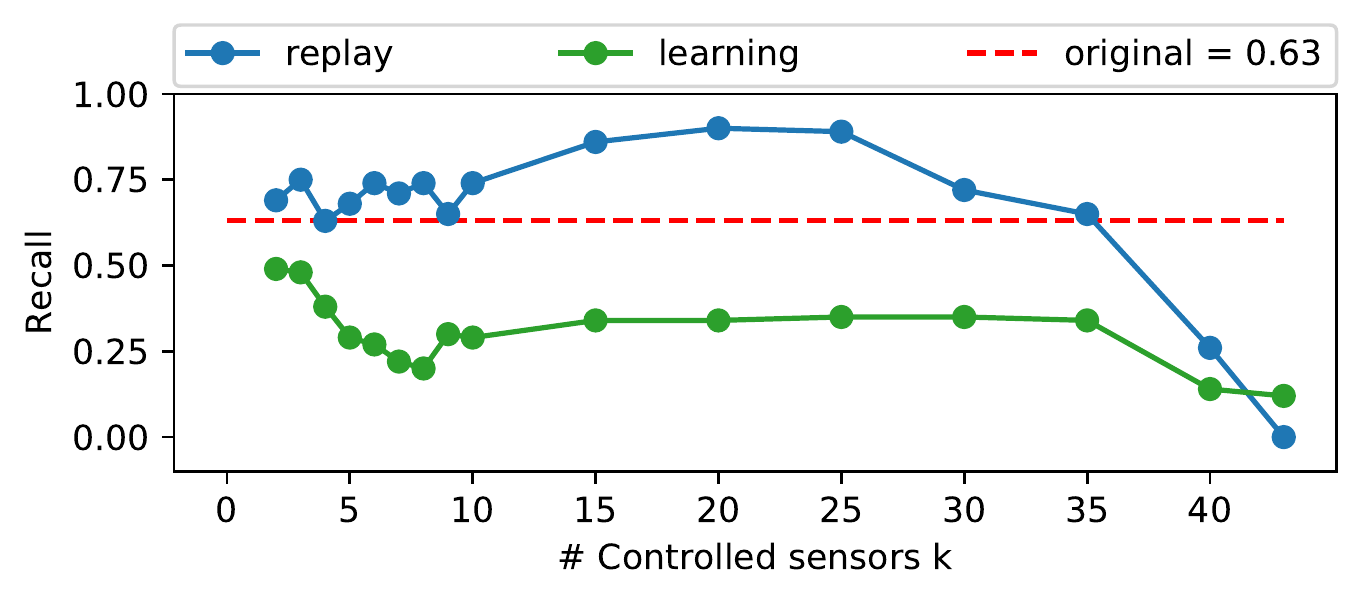}}
    \subfigure[CNN]{\includegraphics[scale=0.55]{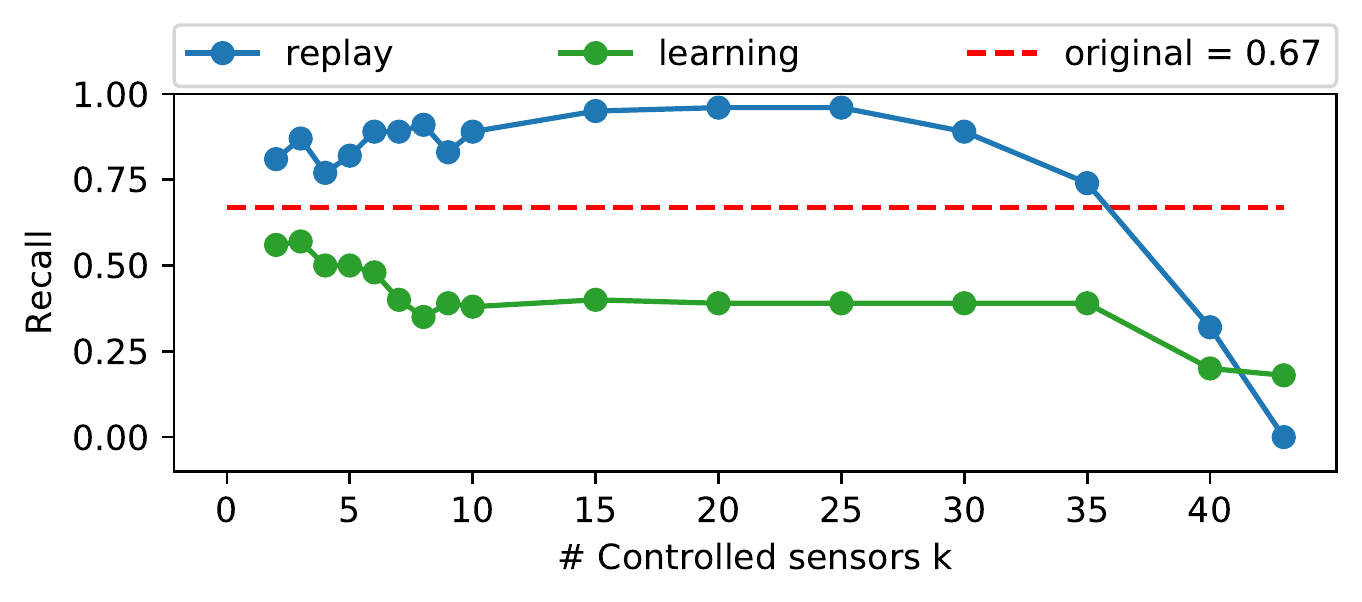}}
    \caption{Generizability of our proposed Learning based attack compared with replay attack. Attack to LSTM (a) and CNN (b) based defenses on BATADAL dataset.}
    \label{fig:generizability}
\vspace*{-2pt}\end{figure*}

\Paragraphs{Data Constrained Attacker,} $(\mathcal{\hat{D}}, \mathcal{X})$.
We also investigated the impact of less available normal data (i.e., a fraction of $\hat{\mathcal{D}}$) on the achieved reduction in detection Recall for the \blackbox attacker. Due to space constraints, the detailed results are presented in Appendix~\ref{sec:impactD}. For BATADAL, the resulting mean detection Recall ranges from 0.14 for 100\% of $\hat{\mathcal{D}}$ available for AE training to 0.22 for 5\% of $\hat{\mathcal{D}}$ available. For WADI, we found that the attacker can leverage on 5\% of normal operations (i.e.,~only 16 hours) to decrease detector Recall to 0.24. 

\Paragraph{Summary of Constrained Attacks findings}
Our results demonstrate that replay attacks perform worse if a limited set of sensors can be manipulated. In particular, if the replay attack is constrained to manipulate less the 95\% of the features the detector's Recall \emph{increases} due to contextual anomalies that are created. For our \approachI and \approachII approaches, this effect does not occur, as the two attacks reduce the detector's Recall without introducing contextual anomalies. In the WADI, the attacker that controls 4\% of the features succeeds in the evasion using the iterative attack.  In addition, an attacker that collects a little amount of data $\hat{\mathcal{D}}$ can train the adversarial autoencoder and perform the \blackbox attack.

\subsection{Generalizability of Learning Based Attack}\label{sec:generalize}

In this section, we evaluate the generalizability of our proposed \approachII{} attack. We consider different reconstruction-based anomaly detectors trained on BATADAL dataset and apply the concealment attack computed with the adversarially trained autoencoder as proposed in our \approachII{} attack. Our target Deep Architecture is an LSTM reconstruction-based anomaly detector as proposed in~\cite{goh2017anomaly} and a CNN reconstruction-based anomaly detector as proposed in~\cite{kravchik2018detecting}. Since those detectors are not available as open source, we implemented their architecture with Keras following the details found in the paper. The LSTM detector was trained to minimize the MSE loss. The network takes 8 timesteps of the multivariate time series as input (input size $8x43$). The input is processed by one LSTM layer with output size $43$, followed by a fully connected layer with $43$ neurons as output dimension. Performance of the LSTM based anomaly detector resulted in \emph{Accuracy}~=~$0.94$, \emph{Precision}~=~$0.89$, \emph{Recall}~=~$0.63$, \emph{FPR}~=~$0.01$. The CNN detector was trained to minimize the MSE loss. The network takes 2 timesteps of the multivariate time series as input (input size $2x43$). The input is processed by three stacked 1D convolutional layers (respectively with $64$, $128$ and $256$ neurons), each convolutional layer is followed by a 1D Max Pooling Layer layer. The last Pooling layer is followed by a flatten and a dropout layer. Dropout layer connects to the fully connected output layer with dimension $43$. CNN based performance after training resulted in \emph{Accuracy}~=~$0.95$, \emph{Precision}~=~$0.90$, \emph{Recall}~=~$0.67$, \emph{FPR}~=~$0.01$. 

Results in Figure~\ref{fig:generizability} show how the detection Recall diminishes when targeted with replay and \approachII{} attacks. In particular, a replay attack can evade detection only in the case in which at least 40 out of 43 sensors are controlled by the attacker (following previous results). Results over \approachII{} attack, despite the different architectures for the offense (Autoencoder) and defense (LSTM and CNN), show that the \approachII{} concealment attack is transferable. In particular, the LSTM architecture appears more vulnerable to concealment attacks, since the \approachII{} attack is achieving higher concealment efficacy than AE and CNN based defenses. Concealment efficacy over CNN defense is comparable to Autoencoder defense, notwithstanding the original Recall of CNN detector is higher than the other considered defense. These results allow us to conclude that the proposed \approachII{} attack efficacy is independent from the Reconstruction-based anomaly detector.

\subsection{Real-time Concealment Attacks} 
\label{sec:real-time_res}
In order to test the real-time feasibility of our attacks, we deployed the anomaly detector~\cite{taormina2018deep} in \wadi, and then attacked it in real-time. We collected 15 hours of normal operations occurring over the ICS. We recorded 62 sensors sampled every 10 seconds.

In this case, we tuned the \emph{window} parameter to 30, which means that the detector is considering the sensor readings occurring in the last 5 minutes. First, we tested the reliability of the system; we left the system running 7 hours without anomalies occurring. We obtained 2 false positives instances occurring for 10 minutes each. We then performed actuators manipulation in the system. In Table~\ref{tab:realTimeTest} we report the summary of the tested anomalies. We replicated anomalies reported in the WADI dataset.

While anomalies were occurring, we also launched our Unconstrained concealment attacks in \wadi to assess their feasibility and efficacy. We tested both the \approachI and \approachII approach in real-time by simulating the sensors value manipulation done by the attacker, all the instances of the anomalies occurring in the system were successfully misclassified. In Table~\ref{tab:realTimeTest}, the last two columns summarize the results of \approachI and \approachII attacks carried out in real-time. In particular, our \approachII modifications took the same time to compute examples as during the earlier experiments (on average, $5 ms$), which is much faster than the sampling rate in the system. 

\begin{table}
    \small
    \centering
    \caption{Real-time detection of process manipulations in \wadi. We replicated anomalies in WADI dataset.}
    \begin{tabular}{cc|c|cc}
        \toprule
            Attack  
            & Duration & Attack& 
            \multicolumn{2}{c}{Detected concealment} \\ 
            \cmidrule(l){4-5} 
            Identifier 
            & (minutes) &  detected&
            Iterative & Learning\\
            \midrule 
            W1 
            & 22 & \cmark &
            \xmark & \xmark \\
            W7 
            & 4 & \cmark &
            \xmark & \xmark \\
            W8 
            & 10 & \xmark&
            \xmark & \xmark \\
            W9 
            & 1 & \cmark &
            \xmark & \xmark \\
            W14 
            & 2 & \cmark &
            \xmark & \xmark \\
        \bottomrule
    \end{tabular}
    \label{tab:realTimeTest}
\end{table}

\section{Related Work}
\label{sec:related}

We now discuss important related work in the area of anomaly detection in CPS, and evasion attacks on classifiers.

\Paragraph{Anomaly Detection in CPS} Detecting stealthy attacks in CPS through the identification of process-based anomalies without requiring a detailed physical model is an active research topic. Had\v{z}iosmanovi\'{c} et al.~\cite{hadziosmanovic2014acsac} use an autoregressive model on time series extracted from modbus PLC traffic, evaluating their approach on data from two water treatment plants; Krotofil et al.~\cite{krotofil2015asiaccs} use a theoretical information approach to detect sensor spoofing attacks; Aoudi et al.~\cite{aoudi18truth} use model-free techniques rooted on singular spectrum analysis to detect structural changes in the process behavior. More recently, in Autonomous Vehicles setting,  control-based techniques used for anomaly detection such as Control Invariant~\cite{choi2018detecting} and Extended Kalman Filters~\cite{bristeau2010hardware, quinonezsavior}, were found vulnerable to several stealthy attacks~\cite{dash2019out, shenposter, quinonezsavior}.

In addition, various proposals in this space use deep learning techniques (usually by training a learning-based model on data gathered during the normal operation of the process) and statistically comparing the sensor readings with the model's prediction at runtime.
Wickramasinghe et al.~\cite{wickramasinghe2018generalization} provide an overview of how Deep Learning techniques can be used in the context of CPS security. Goh et al.~\cite{goh2017anomaly} propose an architecture to detect anomalies over a water treatment testbed with a Recurrent neural network (LSTM-RNN) used to predict sensor readings, and CUSUM to compute the difference between the predicted outputs and the actual sensor readings. Starting from this approach and using the same dataset for evaluation, Kravchik et al.\cite{kravchik2018detecting} suggest the use of a convolutional neural network to perform one-step prediction, while Taormina et al.~\cite{taormina2018deep} propose the autoencoder-based detector (the target of our attacks). With respect to this former category of deep learning-based detectors, our proposed approach is the first to propose a systematic constrained attacker model and identifies vulnerabilities that  affect detector performance. This enables an attacker to hide the physical anomalies induced on the system, that would be otherwise detected. Our experiments show how those attacks can be applied in constrained settings and in real time, making the prior work anomaly detectors ineffective.

\begin{table}[t]
\caption{Recent adversarial learning techniques for evasion, according to the attacker's knowledge and the domain of application. The setting for our attacks is marked with $\star.$}\label{table:algorithms}
\resizebox{\linewidth}{!}{
\begin{tabular}{lccccc}
\toprule
& White Box & Grey Box & \multicolumn{2}{c}{Black Box} \\
\midrule
& $(\mathcal{D}, \mathcal{X}, f, w)$ & $(\hat{\mathcal{D}}, \mathcal{X}, f, \hat{w})$ & \multicolumn{2}{c}{$(\hat{\mathcal{D}}, \hat{\mathcal{X}}, \hat{f}, \hat{w})$} \\
& $\quad$ & $\quad$ & oracle & samples \\
\hline
Malware & \cite{rndic14sp} & \cite{grosse17adversarial} & \cite{xu2016automatically, dang2017evading} & -\\
Image & \cite{carlini17sp, goodfellow14, sharif16ccs}  & \cite{papernot16eurosp} & \cite{papernot17asiaccs} & -\\
ICS & $\star$ & - & - & $\star$ \\
\bottomrule
\end{tabular}}
\end{table}

\Paragraph{Adversarial Learning for Classifier Evasion} The effectiveness of Adversarial Machine Learning to evade ML-based classifiers has been demonstrated in a wide range of applications, ranging from face recognition~\cite{sharif16ccs} to voice recognition~\cite{zhang2017dolphinattack} and malware detection~\cite{xu2016automatically}. Table~\ref{table:algorithms} classifies recent techniques in this domain according to the adversary's knowledge on the classifier's algorithm and training dataset.
In the \approachI scenario (i.e.,~the adversary knows the internals of the trained model and the training set entirely), Rndic and Laskov~\cite{rndic14sp} present a case study on the evasion of PDFRate, a malicious PDF detector based on random forests, using an
\whitebox gradient-based evasion method~\cite{biggio13evasion},  comparing it to a \blackbox mimicry attack, and discussing the attack effectiveness according to different attacker models. After the seminal paper that demonstrated the existence of adversarial examples for neural networks~\cite{szegedy13intriguing}, work has shifted to Deep Learning. Goodfellow et al.~\cite{goodfellow14} study the cause of adversarial examples and devise a fast gradient method to perform adversarial perturbations, demonstrating their results in the image classification context under a perfect-knowledge \approachI scenario. More recently, Carlini and Wagner~\cite{carlini17sp} defeat a defensive technique known as defensive distillation~\cite{papernot16sp}. White box techniques have also been applied to defeat face recognition, also through perturbations in physical objects~\cite{sharif16ccs}.

In more restrictive scenarios, the adversary is only aware of the general structure of the model and how features are extracted. Papernot et al.~\cite{papernot16eurosp} use this imperfect knowledge to build a surrogate model and demonstrate the effectiveness in source-target misclassification (image recognition). Grosse et al.~\cite{grosse17adversarial} generalize the adversarial example crafting algorithm presented in~\cite{papernot16eurosp} to malware detection systems. In other cases, the adversary attacks a classifier while querying the system under attack as an oracle. This is the case of attacks against proprietary online learning systems: to evade an online malware classifier, Xu et al.~\cite{xu2016automatically} leverage the fact that the target systems output the classification score to build a genetic algorithm that morphs the adversarial examples into being undetected. More recently, Dang et al.~\cite{dang2017evading} lifted the assumption of knowing the classification score, attacking oracle-like \blackbox classifiers that only output a binary label; Papernot et al.~\cite{papernot17asiaccs} work similarly in the context of multi-class image classification.

W.r.t. prior work, in our \approachII attack the adversary does not rely on querying the classifier as an oracle, or on building a surrogate learner; instead, we exploit the characteristics of the CPS domain to lift this requirement.

\section{Conclusions}\label{sec:conclusions}

In this work, we started by formalizing an attacker model for real-world ICS and provided AML taxonomy for it. We then presented the first real-time concealment attacks on reconstruction-based anomaly detectors in the context of Industrial Control Systems. We argued that such attacks present four unique challenges, %(R1: Mean-Squared Error loss, R2: Temporal and spatial correlations, R3: Real-time evasion and R4: Feature constrained attacker)
and addressed them proposing \approachI and \approachII attacks. Our \whitebox attacker uses the iterative approach with a detection oracle, while the \blackbox attacker uses an autoencoder to hide anomalies. 

Using data from two water distribution systems, we demonstrated that our attacks are feasible in general, and outperform replay attacks when the attacker is constrained to control of less the 95\% of the features. Moreover, we show that for the BATADAL dataset, our novel \approachII attack using autoencoder was able to reduce detection Recall as efficiently as the \approachI attack (Recall dropped from 0.60 to 0.14 in both cases). Our results demonstrate that the proposed autoencoder based attack achieves successful concealment without knowledge of the targeted Reconstruction-based anomaly detector (only using normal operational data), without knowledge of the physical model equations and is computationally cheap (after training). 

We implemented our attacks in a real testbed and showed that malicious data could be generated on-the-fly, i.e., in between each sampling step (every 10s, actual example generation took on average 5ms for \approachI). That demonstrates that the proposed attacks are allowing attackers to perform constrained concealment attacks on dynamic systems in real-time. In prior work, manipulations are usually performed offline against a dataset or assume that data to be manipulated can be precisely predicted.
Our results show that reconstruction-based attack detectors proposed in prior work are vulnerable to manipulation despite the unique challenges in this setting, and such attacks need to be considered when designing future attack detection schemes.  Implementation is available at our Online Repository\footnote{\url{https://github.com/scy-phy/ICS-Evasion-Attacks}}.

%Acknowledgments 
\begin{acks}
Several authors were supported by the National Research Foundation (NRF), Singapore, under its National Cybersecurity R\&D Programme (Award No. NRF2014NCR-NCR001-40). 
Politecnico di Milano received funding for this project from the European Union's Horizon 2020 research and innovation programme under the Marie Sk\l{}odowska-Curie grant agreement nr. 690972.
\end{acks}

\bibliography{paper}
\bibliographystyle{ACM-Reference-Format}
% Appendix
\appendix

\section{Details of Iterative attack}
\label{sec:iterativeattackdetails}
The attacker essentially has access to an \emph{oracle} of the Deep Architecture, where the attacker can provide arbitrary $\vec{x}$ features and gets the individual values of the  reconstruction error vector $\vec{e}$. The attacker then computes $\max_i\vec{e}$ and finds the sensor reading $r_i$ with the highest reconstruction error from $\vec{x}$. In order to satisfy $\varepsilon(\vec{e'}) < \theta$, the attacker attempts to decrease the reconstruction error $d_i$ error by changing $r_i$. Sensor readings $r_i$ are modified in the range of normal operating values; this guides the computation to a solution that is consistent with the physical process learned by the detector. For example, if normal operations of sensor $r_i$ are in the range $[0, 5]$, the attacker tries to substitute the corresponding value of $r_i$ according to its range to see if the related reconstruction error decreases. This results in $\vec{x}' = [r_1, \dots, r_i',\dots, r_n]$, where $d_i' < d_i$ and, accordingly, $\varepsilon(\vec{e}' ) < \varepsilon(\vec{e})$.
Algorithm~\ref{alg:WEM} is the pseudo-code applied to compute sensor readings modifications.

In order to find the value of $r_i$ that decreases $\varepsilon(\vec{e})$ the most, we can introduce $X$ as the matrix containing the mutations of $\vec{x}$  w.r.t. $r_i$.
\[
X =
\begin{bmatrix}
      r_1&\dots & r_i^1&...& r_{n} \\
      r_1& \dots &r_i^2&...& r_{n} \\
      \vdots & \ddots & \vdots & \ddots & \vdots \\
      r_1& \dots & r_i^m&...& r_{n}
\end{bmatrix}
\]
were $r_i^k \in {\text{\emph{normal operations values for senor i}}}$. Among the all mutations, we select the one that generates the lower reconstruction error $\varepsilon(\vec{e})$. After choosing the best value over the variable $r_i$ the algorithm repeats until a solution with average reconstruction error lower than $\theta$ is found. 

Two stopping criteria are put in place: \emph{patience} and  \emph{budget}. It could happen that no lower reconstruction errors $d_i$ are found by changing the value of a chosen reading $r_i$. In this case, we try to change the other readings in descending order of reconstruction error. \emph{patience} mechanism is put in place to avoid wasting of computation. If no improved solutions are found in \emph{patience} iterations, the input is no more optimized.

According to the communication mechanism between PLCs and SCADA, the attacker may be constrained to send the data in a certain amount of time. \emph{budget} is the maximum amount of times that loop at Line~\ref{lst:line:loop} (Algorithm~\ref{alg:WEM}) can be performed. After \emph{budget} attempts without finding a set of modified readings that satisfies $\varepsilon(\vec{e}') < \theta$, the input is no more optimized, and no solution is found. 

Exiting the loop at Line~\ref{lst:line:loop} due to a stopping criterion is not providing a misclassified example. Even though a solution such that  $\varepsilon(\vec{e}') < \theta$ is not found, the resulting tuple is likely to have a lower $\varepsilon(\vec{e})$, i.e.,  $\varepsilon(\vec{e}) > \varepsilon(\vec{e}') > \theta$.

\begin{algorithm}[tb]
\caption{White Box concealment attack}\label{alg:WEM}
\begin{algorithmic}[1]
\Procedure{Conceal}{$\vec{x}$}
\State $c \gets 0$\Comment{number of changes}
\State $i \gets 0$\Comment{last optimization}
\State \textit{solved} $\gets$ \textit{False}
\State $\vec{e}\gets$  \textit{compute\_reconstruction\_errors(} $\vec{x}$ \textit{)}
\State  \textit{previous\_best\_error} $\gets  \varepsilon(\vec{e})$
\Comment{access oracle}
\State $\vec{e}\gets$  \textit{sort\_descending(}$\vec{e})$
\While{$!(solved) \ \AND\  (c-i) < patience \ \AND\ c < budget$} \label{lst:line:loop}
	\State  \textit{f} $\gets$  \textit{choose\_feature\_to\_optimize (}$\vec{e})$
	\State $X \gets$ \textit{compute\_matrix\_of\_mutations}$(\vec{x}, f)$ \label{lst:line:matrix}
	\State $x', \vec{e}' \gets $\textit{find\_best\_mutation}$(\vec{X})$ \label{lst:line:matrix_comp}	
	\If{$ \varepsilon(\vec{e}') < previous\_best\_error$}
		\State \textit{previous\_best\_error} $\gets \varepsilon(\vec{e}')$
		\State \textit{new\_best} $\gets \vec{x}'$
	\Else
		\State $i \gets c$
	\EndIf
	\If{$\varepsilon(\vec{e}') < \theta$}
		\State $solved \gets True$
	\EndIf
    \State $c \gets c + 1$
	\State $\vec{e}\gets $\textit{sort\_descending}$(\vec{e}')$
\EndWhile
\State \textbf{return} \textit{new\_best}
\EndProcedure
\end{algorithmic}
\end{algorithm}

\section{Definition of Constraints}
\label{sec:constraintsdef}
In order to study the impact of this best-case constraints, we selected $k$ features for every attack that can be modified. Then we studied how replay, \approachI, and \approachII attacks perform when these constraint are applied. We defined the constrains as follows: starting from the results of \approachI and \approachII attacks we determined the $k$ features that were changed most frequently (over the course of each attack). The intuition behind this is as follows: features that are modified most often in the unconstrained case are assumed to have the highest impact on the performance of the attack. We are assuming a best case scenario for the attacker, in which she was able to choose  the $k$ our of $n$ features to maximise her efficiency. Then, we created 11 sets of $k$ features that can be modified by the attacker (with different counts $k$ of features to be manipulated, with the maximal number determined by the dataset used). In the case of \approachI and \approachII attack, we limited the adversarial example exploration to the $k$ features extracted for the considered approach. Effectively, that implies that we only used the allowed $k$ features out of the \approachII model, while the \approachI model was able to learn modifications to the $k$ allowed features that would minimize the detector accuracy. 
In the case of replay attack we applied the same replay strategy introduced before but we replayed only the selected $k$ features extracted from the \approachI approach. We note that this choice (replay the features extracted from the \approachI approach) was made to reflect worst case scenario, i.e., an attacker that is able to replay exactly the $k$ features that an \approachI attacker would replay.

\balance
\section{Learning Based attack: impact of $\mathcal{D}$ dimension}
\label{sec:impactD}
Another aspect that we investigated is the impact of $\hat{\mathcal{D}}$ on the applicability of \approachII attack. Especially, we are interested in understanding how much normal operational data the attacker needs to conduct the proposed  \approachII attack. We investigated the impact of less available normal data (i.e., a fraction of $\hat{\mathcal{D}}$) on the achieved reduction in detection Recall for the \approachII attacker. We performed a sensitivity analysis by random sampling normal operations data 10 times for each one of the considered percentages of data. Then, we trained an adversarial network for each sampling of the data percentage (50 adversarial networks trained for each dataset). As result we computed the sample mean (\smean) and sample standard deviation (\sstd) of the resulted detection Recall by using the different \approachII networks. 
\begin{table*}
    \small
	\centering
	\caption{Impact of fraction of $\hat{\mathcal{D}}$  on concealing capacity.}
		\begin{tabular}{ccccccccccccc}
			\toprule
			&  & \multicolumn{11}{c}{Recall for Black Box \% of $\hat{\mathcal{D}}$} \\
			\cmidrule(l){3-13}
			& Original & 100\% & \multicolumn{2}{c}{75\%} & \multicolumn{2}{c}{50\%} & \multicolumn{2}{c}{25\%} & \multicolumn{2}{c}{10\%} & \multicolumn{2}{c}{5\%} \\
			\cmidrule(l){3-3} \cmidrule(l){4-5} \cmidrule(l){6-7} \cmidrule(l){8-9} \cmidrule(l){10-11} \cmidrule(l){12-13}
			Data& Recall& & \smean & \sstd & \smean & \sstd & \smean & \sstd & \smean & \sstd & \smean & \sstd \\
			\midrule
    	    B & $ {0.60} $ & $ {0.14} $ & $ {0.15} $ & $ {0.02} $ & $ {0.15} $ & $ {0.02} $ & $ {0.16} $ & $ {0.03} $ & $ {0.25} $ & $ {0.09} $ & $ {0.22} $ & $ {0.09} $\\
    		W & $ {0.68} $ & $ {0.31} $ & $ {0.27} $ & $ {0.06} $ & $ {0.26} $ & $ {0.01} $ & $ {0.24} $ & $ {0.03} $ & $ {0.24} $ & $ {0.04} $ & $ {0.24} $ & $ {0.06} $\\
			\bottomrule
		\end{tabular}
	\label{tab:impactD}
\end{table*}

For BATADAL, the resulting mean detection Recall ranges from 0.14 for 100\% of $\hat{\mathcal{D}}$ available for AE training to 0.22 for 5\% of $\hat{\mathcal{D}}$ available. For WADI, the resulting mean detection Recall ranges from 0.31 for 100\% of $\hat{\mathcal{D}}$ available for AE training to 0.50 for 5\% of $\hat{\mathcal{D}}$ available (compared to 0.68 without concealment).
Results over BATADAL dataset show, performance of the attacker's adversarial network is performing almost the same if trained with 100\% to 25\% of data. Lower than 25\% of the data we notice substantial performance degradation. Looking at standard deviation, we notice that less data (10\%. 5\%) causes high model variance. To perform the \approachII attack the attacker needs 25\% of data to guarantee evasion success.

In the case of WADI dataset performance of the adversarial network increases diminishing the number of data available to the attacker, this means that with less data the attacker's Autoencoder generalizes better. WADI water distribution network is small and the three stage are repetitive. Information contained in 5\% of the data (16 hours of recordings) could be enough to model the system behavior.

\section{Discussion}
\label{sec:discussion}
We showed that replay attacks (while not requiring machine learning algorithms) is only efficient when all sensor readings replayed. Thus, replay attacks do not represent a viable solution for hiding anomalies when the attacker can act on a limited set of sensor readings. In particular, replay attacks introduce contextual anomalies since sensor readings will not be consistent any longer. 

\begin{figure}[tb]
    \centering

    \includegraphics[width=0.99\linewidth]{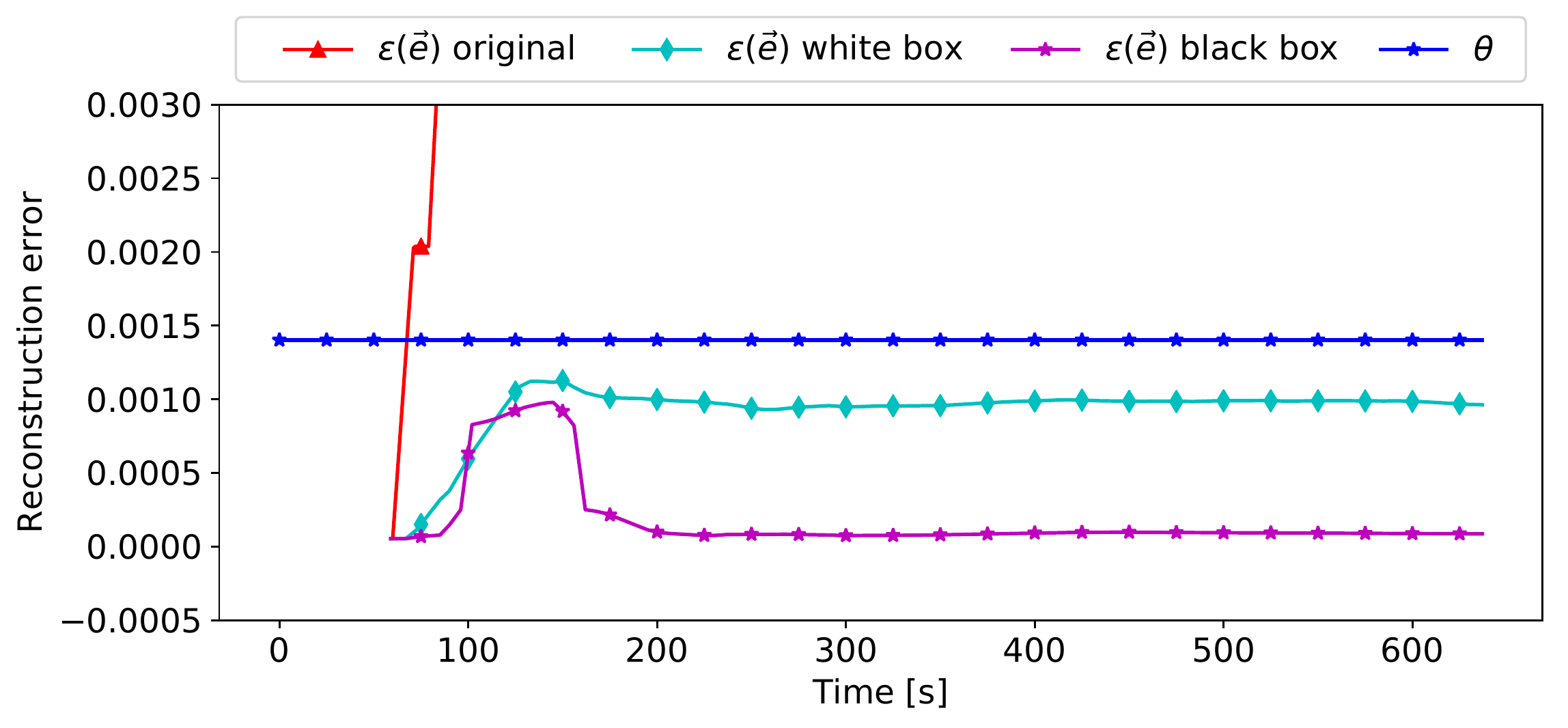}
    \caption{Comparison of concealment results. While the Recall after concealment in both white and \approachII goes to $0$, we can see how the two approaches are behaving differently. We plot the average reconstruction error over time ($\varepsilon(\vec{e})$) and the threshold $\theta$. }
    \label{fig:concealment_comparison}
\end{figure}

We now discuss the quality of results coming from the proposed approaches. Figure~\ref{fig:concealment_comparison}, represents the comparison between trend of $\varepsilon(\vec{e})$ wrt. the threshold $\theta$ during the whole actuators' manipulation done in one attack from WADI dataset. Comparing the white and \approachII $\varepsilon(\vec{e})$ results, we  notice that the solution provided by the \approachI algorithm is closer to $\theta$ than the \approachII solution. This is because the \approachI algorithm is looking at $\theta$ value to decide whether to stop the computation. Black box is not performing any optimization wrt. the attacked detector, so it is providing a solution that is matching the learned physical behavior, and what the detector expects from a non-anomalous sample. After second 200, the magenta line shows that the $\varepsilon(\vec{e})$ is around $0$, meaning that we are sending inputs that are in line with the detector expected behavior.

\begin{table*}
\setlength{\tabcolsep}{5pt}
\renewcommand{\arraystretch}{.75}
\small
\caption{Impact of Partially Constrained attacker (best-case scenario and topology based scenario). By decreasing the number of features that the attacker can control, we  notice that replay attack performance decreases drastically. This is due to contextual anomalies introduced as part of large-scale replay, while both black and \approachI approaches avoid this problem.}
\label{tab:BATADAL-WADI-constrained_result}
	\centering
			\resizebox{\linewidth}{!}{
\begin{tabular}{cccccccccccccccccccc}
			\toprule
			& Original & &\multicolumn{16}{c}{Recall vs. \# of Controlled sensors $k$ (43 features BATADAL)}  & Topology \\	\cmidrule(l){4-19}
			Data & Recall & Experiment & 
			43 &    40 &    35 &    30 &    25 &    20 &    15 &    10 &    9  &    8  &    7  &    6  &    5  &    4  &    3  &    2  &1 PLC\\ \midrule
			\multirow{3}{*}{B} & \multirow{3}{*}{$0.60$} & replay  & 0 &  0.25 &  0.69 &  0.83 &  0.94 &  0.93 &   0.9 &  0.78 &  0.69 &  0.79 &  0.77 &  0.77 &  0.71 &  0.68 &  0.78 &  0.73 & 0.36\\
			&&\approachI &  0.14 &  0.17 & 0.17 &  0.15 &  0.16 &  0.22 &  0.22 &  0.22 &  0.22 &  0.25 &  0.25 &  0.25 &  0.27 &  0.35 & 0.51 &  0.52 &0.34 \\
			&&learning &  0.14 &  0.16 &  0.35 &  0.37 &  0.37 &  0.38 &  0.38 &  0.35 & 0.35 &  0.24 &  0.27 &  0.32 &  0.34 &  0.49 &  0.56 &  0.55 & 0.34\\
            \bottomrule \\
	    \end{tabular}}
		\resizebox{\linewidth}{!}{
\begin{tabular}{cccccccccccccccccccccc} 
			\toprule
			& Original & &\multicolumn{18}{c}{Recall vs. \# of Controlled sensors $k$ (82 features WADI)} & Topology\\	\cmidrule(l){4-21}
			Data & Recall & Experiment & 
			82 &    80 &    70 &    60 &    50 &    40 &    30 &    20 &    15 &    10 &    9  &    8  &    7  &    6  &    5  &    4  &    3  &    2   & 1 PLC\\ \midrule
			\multirow{3}{*}{W} & \multirow{3}{*}{$0.68$} &  replay & 0.07 &  0.15 &  0.55 &   0.7 &  0.74 &  0.79 &  0.89 &  0.87 &  0.85 &  0.65 &  0.64 &  0.62 &  0.62 &  0.62 &   0.6 &  0.44 &   0.4 &  0.49 & 0.64\\
			&&\approachI &  0.07 &  0.07 &  0.07 &  0.07 &  0.07 &  0.07 &  0.07 &  0.07 &  0.07 &  0.08 &  0.09 &  0.12 &  0.12 &  0.12 &  0.12 &  0.12 &  0.14 &  0.25 & 0.12\\
            &&learning  &  0.31 &  0.31 &  0.31 &   0.4 &   0.4 &   0.4 &  0.45 &  0.41 &  0.41 &  0.32 &  0.36 &  0.36 &  0.36 &  0.34 &  0.27 &  0.27 &  0.27 &  0.29 &0.36\\
			
            \bottomrule
	    \end{tabular}}
\end{table*}

\begin{table*}[tbh!]
\small
	\centering
	\caption{Generizability of Learning based attack. Attack to LSTM and CNN based defenses on BATADAL dataset.}
	\label{tab:generizability}
			\begin{tabular}{ccccccccccccccccccc}
			\toprule
			& Original & &\multicolumn{16}{c}{Recall vs. \# of Controlled sensors $k$ (43 features BATADAL)} \\	\cmidrule(l){4-19}
			Data & Recall & Experiment & 
			43 &    40 &    35 &    30 &    25 &    20 &    15 &    10 &    9  &    8  &    7  &    6  &    5  &    4  &    3  &    2  \\ \midrule
			\multirow{2}{*}{LSTM} & \multirow{2}{*}{$0.63$} &  replay   &     0 &  0.26 &  0.65 &  0.72 &  0.89 &   0.9 &  0.86 &  0.74 &  0.65 &  0.74 &  0.71 &  0.74 &  0.68 &  0.63 &  0.75 &  0.69 \\
			&&learning &  0.12 &  0.14 &  0.34 &  0.35 &  0.35 &  0.34 &  0.34 &  0.29 &   0.3 &   0.2 &  0.22 &  0.27 &  0.29 &  0.38 &  0.48 &  0.49 \\ \hline
			\multirow{2}{*}{CNN} & \multirow{2}{*}{$0.67$} & replay   &   0 &  0.32 &  0.74 &  0.89 &  0.96 &  0.96 &  0.95 &  0.89 &  0.83 &  0.91 &  0.89 &  0.89 &  0.82 &  0.77 &  0.87 &  0.81 \\
			&&learning &  0.18 &   0.2 &  0.39 &  0.39 &  0.39 &  0.39 &   0.4 &  0.38 &  0.39 &  0.35 &   0.4 &  0.48 &   0.5 &   0.5 &  0.57 &  0.56 \\
			\bottomrule
 \end{tabular}
\end{table*}

\end{document}